\documentclass[fleqn,usenatbib]{mnras}

\usepackage{newtxtext,newtxmath}

\usepackage[T1]{fontenc}
\usepackage{ae,aecompl}

\usepackage{graphicx}	
\usepackage{amsmath}	
\usepackage{amssymb}	
\usepackage{xcolor}     
\usepackage{soul}       
\usepackage{url}        

\title[Hadronic emission along AGN jets]{On the relative importance of hadronic emission processes along the jet axis of Active Galactic Nuclei}

\author[Mario R.\ Hoerbe et al.]{
Mario R. Hoerbe,$^{1,2,3}$\thanks{E-mail: mario.hoerbe@ruhr-uni-bochum.de}
Paul J.\ Morris,$^{2,4}$
Garret Cotter,$^{2}$
Julia Becker Tjus$^{1,3}$
\\
$^{1}$Ruhr-Universit\"at Bochum, Theoretische Physik IV: Plasma-Astroteilchenphysik, Universit\"atsstrasse 150, 44801 Bochum, Germany\\
$^{2}$University of Oxford, Oxford Astrophysics, Denys Wilkinson Building, Keble Road, Oxford, OX1 3RH, United Kingdom\\
$^{3}$Ruhr Astroparticle and Plasma Physics Center (RAPP Center), Ruhr-Universit\"at Bochum, 44780 Bochum, Germany\\
$^{4}$DESY, Platanenallee 6, D-15738 Zeuthen, Germany\\
}

\date{Accepted 2020 June 1 for publication in MNRAS}

\pubyear{2020}

\begin{document}
\label{firstpage}
\pagerange{\pageref{firstpage}--\pageref{lastpage}}
\maketitle

\begin{abstract}
With the coincident detection of a gamma-ray flare and a neutrino from the blazar TXS 0506+056, Active Galactic Nuclei (AGN) have been put into focus as possible sources of the diffuse neutrino flux. We present a space and time-resolved model of the high-energy particle emission of a plasmoid assumed to travel along the axis of an AGN jet at relativistic speed. This was achieved by modifying the publicly available {\small CRPROPA} (version 3.1+) propagation framework which in our work is capable of being applied to source physics on sub-kpc scales. The propagation of a population of primary protons is modelled in a purely turbulent magnetic field and we take into account interactions of these protons with photons scattered from the accretion disc, synchrotron radiation emitted by ambient relativistic electrons, as well with themselves and with other ambient matter.
Our model produces a PeV-neutrino flare caused mainly by photo-hadronic interactions of primaries with the accretion disc field. Secondary high-energy gamma-rays partly attenuate with the ambient photon fields whose combined optical depths achieve their minimal opacity for photons of around 10 TeV. Thus, our model is well capable of producing neutrino flares with a significantly reduced emission of gamma-rays in jets with a hadronic jet component which in the future can be fit to specific AGN flare scenarios.

\end{abstract}

\begin{keywords}
astroparticle physics -- galaxies: jets -- methods: numerical -- neutrinos -- gamma-rays: galaxies
\end{keywords}



\section{Introduction}

Multi-messenger astrophysics has substantially contributed to our understanding of astrophysical phenomena. The approach of viewing sources in their full picture ranging from hadronic cosmic rays, radio-waves up to gamma-rays, high-energy neutrinos and gravitational waves allows for previously unprecedented insights into the correlations between high-energy astrophysical processes. Such processes may to occur in active galactic nuclei (AGN) which are the most luminous, continuous sources of radiation with luminosities in excess of $10^{37}~\rm{W}$ ($10^{44}~\rm{erg}~\rm{s}^{-1}$) \citep{becker2008high_NeutrinoReview} and are therefore candidates to accelerate cosmic rays towards the highest energies and so may contribute to the diffuse astrophysical high-energy neutrino flux as measured by IceCube at Earth \citep{icecube2013evidence_for_HENu}. In particular, blazars are known for their high emission variability which implies the existence of highly energetic processes in their jets which due to their orientation get Doppler-boosted into Earth's rest frame. In literature, blazar jets are believed to be dominated by leptonic particle processes which well fit observational quiescent data of blazar spectral energy densities (SEDs) (\citet{reynolds1996matter_leptonJet, wardle1998electron_leptonJet, potter2012synchrotron, potter2015new_jetModel}). However, (blazar) jets have been speculated to also contain hadronic components which would not only contribute to the emission of electromagnetic radiation in blazars (\citet{bottcher2013leptonic_inferHadronsInEMemission}) but also lead to the production of secondary high-energy neutrinos and gamma-rays (\citet[e.g.]{becker2005diffuse,bb2009, murase2012blazars_blazarsGenerateUHECR, tjus2014high_HENuRadioGalaxies}). This assumption is supported by general evidence of up to PeV high-energy-neutrinos of astrophysical origin observed by \citet{icecube2013evidence_for_HENu} of which a neutrino of $290~\rm{TeV}$ could be associated with the flare of the gamma-ray blazar TXS 0506+056 with a significance of $3\sigma$ by combining data of the IceCube neutrino observatory and Fermi-LAT (\citet{icecube2018multimessenger_PKS_NEHuGamma}). This detection initiated a dedicated search for neutrino clustering in the $\sim 10$~years of IceCube data around the position of TXS 0506+056, with the result that there was an enhanced flux of neutrinos in a half-year period from September 2014 to March 2015, with a statistical significant of $\sim 3\,\sigma$ of being incompatible with the background hypothesis. It is interesting to note that this potential flare is not accompanied with a gamma-ray flare.

Models of high-energy neutrino and electromagnetic up to gamma-ray emission in jets of AGN cover a variety of scenarios and parameter spaces. Blazars are known to be highly variable across the electromagnetic spectrum, with a variety of models put forward to explain these flares. Such models include, among others, particle acceleration via internal shocks in the jet \citet{becker2005diffuse, eichmann2010plasmoidLeptonic} and reconnection driven plasmoids \citet{giannios2013reconnection_PlasmoidGamma, morris2019feasibility_paulsPlasmoid}. Such relativistic and compact structures have been discussed in the literature since the 1960s \citep{rees1966,vanderlaan1966}. What we refer to as \textit{plasmoid} is often called \textit{blob} in the literature, meaning compact, dense structures traveling with relativistic speeds along the jet axis. The term \textit{plasmoid} is preferred here, as it is typically used in the context of the plasmoid creation via magnetic reconnection events. In this scenario, the injection of a relativistic plasma into the system (here the AGN jet) can lead to reconnection events that under certain circumstances lead to the plasmoid instability that breaks down the streaming plasma into small \textit{blobs}, i.e.\ the plasmoids. In this scenario, charged particles can be accelerated in the reconnection events. While non-relativistic reconnection is limited to below knee energies \citep{lyutikov-komissarov-sironi-porth-2018}, relativistic reconnection can be much more efficient \citep{sironi-spitkovsky-2014} and can even be used to solve the injection-problem by serving as pre-accelerators before the particles enter shock fronts, where they can be accelerated to ultra high-energies.

The modelled hadronic component of solely inelastic proton-proton interaction was discussed in \citet{eichmann2012plasmoidHadronic} up to proton-dominated emission (\citet{sahakyan2018lepto_PKShadronDominated}) and broader multi-messenger modelling of leptonic and (lepto-)hadronic emission in \citet{christie2018radiative_EquilibriumPlasmoids_B1G, keivani2018multimessenger_blazarNuGamma} for the case of PKS 0506+056. Flare models not based on magnetic field dynamics invoke external factors such as e.g.\ gas clouds entering the base of jets (\citet{dar1997hadronic_cloudEntersJet, araudo2010gamma_cloudEntersJet, zacharias2019extended_cloudAblationByJet}) or jets of former binary AGN drilling through their own dust tori after being redirected by the merger of their host black holes (\citet{gergely2009spin_flip}) which all tend to be more hadron-dominated due to the nature of their occurrence.

We aim to investigate the relative importance of hadronic emission channels as a function of distance from the supermassive black hole. We provide space and time-resolved analyses of selected interaction processes and environmental circumstances which are expected to majorly influence the evolution of particle emission. This was achieved by modifying the publicly available {\small CRPROPA} (short for `Cosmic Ray Propagation') framework\footnote{\url{https://github.com/CRPropa/CRPropa3}} which in our work is, as in contrast to previous work of e.g. \citet{zhang2020neutral_CRPropa}, is also capable of being applied to sub-kiloparsec environments such as (but not constrained to) AGN jets, as opposed to (inter-)Galactic propagation scenarios in the original version for which {\small CRPROPA} was designed. The changes we introduced in the described version of {\small CRPROPA} include:
\begin{enumerate}
    \item Photon fields with arbitrary spectral shape and redshift dependence
    \item Interactions of protons with other protons
    \item Optional, arbitrary dependence of all interaction modules, including photon fields, on space and time
    \item Interaction tags of output data to backtrack interaction from which they emerged
\end{enumerate}

The model presented analyses the evolution of a primary, hadronic population of protons with a temporal resolution in the order of $10^2$ seconds while residing in a plasmoid which moves at relativistic velocity along the jet. Due to the explicit simulation of individual particles we are able to also consider effects of the finite extent of the interaction region such as the plasmoid size and geometric scaling of fields along the jet axis. Our model is scalable such that in the future it may be applied to a broader range of emission scenarios as we investigate the relative influence of environmental parameters in the interplay of magnetic- photon- and matter fields on the primary proton population and derive implications on the energies and time-scales secondary particles including neutrino and gamma-ray flares.

This paper is organised as follows: In Sec.~\ref{sec:model} we describe our plasmoid model covering the magnetic field configuration in Sec.~\ref{subsec:magneticField}, possible ambient matter distributions Sec.~\ref{subsec:photonFields}, and dominant photon fields in Sec.~\ref{subsec:ambientMatter}. Details of interactions with these targets may be inferred from Sec.~\ref{subsec:simSetup}. Our main results are summarised in Sec.~\ref{sec:results} based on the parameter values we used (Sec.~\ref{subsec:parameterSetup}). Here, we analyse the relative importance of processes (Sec.~\ref{subsec:importanceOfProcesses}) that govern the evolution of the primary protons but also secondary populations of neutrons, neutrinos and gamma-rays whose space- and time-resolved signatures in Sec.~\ref{subsec:plasmoidEvolution}. We discuss the impact of parameters other than those used in our model in Sec.~\ref{subsec:impactOfParameterVariations} where we refer to a broader range of applications of our model in literature. Sec.~\ref{sec:conclusions} concludes our work.

\section{Model}
\label{sec:model}

We begin by injecting a spectrum of protons and electrons which we assume have been accelerated to highly relativistic energies and reside within an AGN jet. In this work, we consider the radiative processes from a subset of such relativistic plasma which are contained within a relativistically moving plasmoid as e.g.\ used by \citet{giannios2009fast_blobSize_blobProcesses, eichmann2012plasmoidHadronic, christie2018radiative_EquilibriumPlasmoids_B1G, keivani2018multimessenger_blazarNuGamma, morris2019feasibility_paulsPlasmoid}. The jet is assumed to initially be in equipartition among its energetically significant components such that the energy density of the magnetic field (in SI units) $U_b=B^2/2\mu_0$ equals the energy density of all particles which constitute the jet plasma. Equipartition therefore reads as
\begin{equation}
	U_b = U_{\rm{particles}} = U_{\rm{p}} + U_{\rm{e}}
	\label{eq:equipartition}
\end{equation}
in which $U_{p/e}$ denote the energy densities of relativistic populations of protons and electrons. Depending on the exact circumstances of matter injection into the jet and the acceleration mechanisms at hand, the ratio of energy intrinsic to the proton population as opposed to the energy of the electron population may vary. By choosing a free dimensionless parameter $\chi=U_p/U_e$ one may parameterise the distribution of energy among leptonic and hadronic components according to
\begin{equation}
	U_p(\chi)=\frac{U_b\chi}{\chi+1} ~~~\mbox{and}~~~U_e(\chi)=\frac{U_b}{\chi+1}\quad.
	\label{eq:energyBudget}
\end{equation}

\begin{figure*}
	\includegraphics[width=\textwidth]{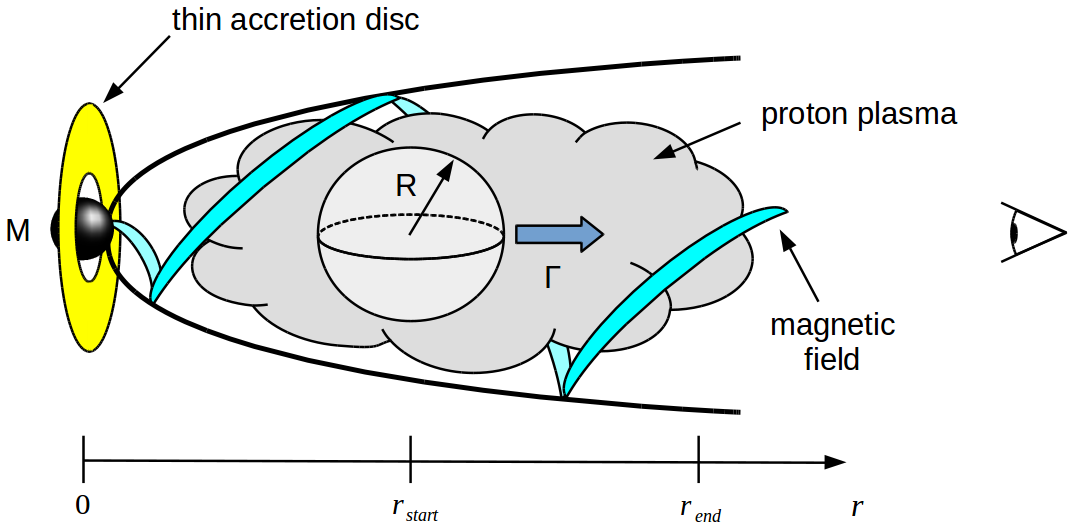}
	\caption{Illustration of the physical model used in this work. Inside an AGN jet propagates a plasmoid of radius $R$ at Lorentz factor $\Gamma$ along the jet axis. Particles inside this plasmoid are deflected by magnetic fields and scatter with various photon fields as well as ambient, hadronic plasma.}
	\label{fig:figure1_jetSketch}
\end{figure*}

The object investigated in this work is a flaring region represented by a plasmoid embedded into the proposed AGN jet as shown in Fig.~\ref{fig:figure1_jetSketch}. The plasmoid extends over a spherical region with radius $R$ and travels at a Lorentz factor of $\Gamma$ along the jet axis for a distance from $r_{\rm{start}}$ to $r_{\rm{end}}$. The flaring event itself is assumed to produce relativistic protons inside the plasmoid volume in which a rather general acceleration scenario is assumed to have taken place, as e.g.\ initiated by a propagating shock front inside the jet, leading to chaotic high-energy particle production. In order to keep the model general, an isotropic delta-functional particle injection approach is chosen, i.e. an instantaneous injection of a proton density $n_p$ at a point of time $t_0$ at each point of the entire plasmoid. The energy distribution is modelled to have a power-law of index $\alpha_p$ to reflect Fermi-like particle acceleration backgrounds, i.e. $\mathrm{d}n_p/\mathrm{d}E_p\propto E_p^{-\alpha_p}$. The population of electrons is assumed to have been accelerated in the same process also leading to a power-law energy distribution with spectral index $\alpha_e$ with $\mathrm{d}n_e/\mathrm{d}E_e\propto E_e^{-\alpha_e}$.\\
The relativistic protons encounter three main types of targets inside the plasmoid:
\begin{enumerate}
    \item A turbulent magnetic field deflects both ambient protons and electrons. The magnetic field is turbulent on scales larger than 
    the radius of the plasmoid and quickly isotropises the velocity vectors of both particle populations. 
    \item Electrons being deflected by this magnetic field will emit synchrotron radiation photons that in return may interact with the relativistic protons via photo-hadronic interactions causing the production of secondary gamma-rays and neutrinos. The plasmoid is illuminated by a black body radiation field which we assume to be from the accretion disc. This field is Lorentz transformed into the plasmoid rest frame and leads to further photo-hadronic interactions.
    \item An ambient matter field is assumed to exist and modelled as a proton plasma with the same properties as those in the plasmoid which also propagates along the jet. This matter leads to hadron-hadron interactions producing secondary gamma-rays and neutrinos.
\end{enumerate}
We track the energy losses and particles produced within the rest frame of the plasmoid in which they are initially defined. Detailed descriptions of the components mentioned above can be found in the subsequent sections. 

\subsection{Magnetic Field}
\label{subsec:magneticField}

This work takes the presence of a turbulent magnetic field into account whose correlation length $L_c$ is smaller than the radial extent of the plasmoid itself. This guarantees that a uniform velocity distribution of a test population of charged particles becomes nearly isotropic after propagating a length of $L_c$. The magnetic field strength has got a root mean square value of $B_0$ at the base of the jet (defined at $r_{\rm{start}}$) and changes in strength along the radial extent of the jet as we assume that the total magnetic energy is conserved along the length of the jet. This can be demonstrated by assuming that the jet is nearly conical as shown in \citet{blandford1979relativistic_conicalJet, asada2012structure_jetParabolicToConical} with an opening angle $\theta$ and radius $R_{\rm{jet}}$ which increases along the jet axis according to $R_{\rm{jet}}(r) \propto r\tan\theta$. A slice of the jet at position $r$ of thickness $\mathrm{d}r$ is then geometrically described by a truncated cone with volume element $\rm{d}V_s$
\begin{equation}
    \begin{array}{rl}
    	\mathrm{d}V_s(r)=&\frac{\pi}{3}\left(R_{\rm{jet}}^2(r) + R_{\rm{jet}}(r)R_{\rm{jet}}(r+\mathrm{d}r) + R_{\rm{jet}}^2(r+\mathrm{d}r)\right)\mathrm{d}r\\
		\propto & \left(r^2 + r(r+\mathrm{d}r) + (r+\mathrm{d}r)^2\right)\mathrm{d}r	\propto r^2~~~.
	\end{array}
    \label{eq:jetSliceVolume}
\end{equation}
If the magnetic field possesses an initial energy density of $U_b$, the slice contains a magnetic energy $\mathrm{d}E_b$ of
\begin{equation}
	\mathrm{d}E_b = U_b(r)~\mathrm{d}V_{\rm{slice}} \propto B^2(r)\cdot r^2 ~~~.
\end{equation} 
Since we assume that the magnetic field itself does not suffer any energy losses along the jet, there has to be the same energy content in each slice of the jet regardless of position $r$ i.e.\ $\mathrm{d}E_b = const. \Rightarrow \mathrm{d}E_b(r_1)=\mathrm{d}E_b(r_2)$. Thus, it follows that the magnetic field itself has to scale according to $r^{-1}$ in order to conserve energy which we model as e.g.\ done in \citet{blandford1979relativistic_conicalJet, potter2012synchrotron}
\begin{equation}
	B(r) = \frac{\kappa_B}{r}
	\label{eq:BfieldScaling}
\end{equation}
using a normalization constant $\kappa_B$. Note that the quantity $B_0$ is chosen to approximately describe a characteristic magnetic field strength at $r_{\rm{start}}$ in order to determine the global scaling of the magnetic field along the jet axis. Importantly, the magnetic field itself still is turbulent on length scales of the plasmoid radius yet is subject to an overall decrease in field strength proportional to $r^{-1}$.

\subsection{Ambient Matter}
\label{subsec:ambientMatter}

Other matter that is also being injected into the jet is modelled as a proton plasma with an average energy equal to a mean energy of $E_{\rm{plasma}}$, as we assume collisions occur between protons contained within the plasmoid and/or other protons that reside inside the jet. In analogy to the argumentation of energy conservation in the case of the scaling function of the magnetic field strength, one can impose a scaling function on the number density of the ambient proton plasma according to
\begin{equation}
	n_{\rm{plasma}}(r_1)\cdot r_1^2 = n_{\rm{plasma}}(r_2)\cdot r_2^2
	\label{eq:ambientMatterFieldScaling}
\end{equation}
from which follows that the proton plasma density itself has to scale according to $r^{-2}$ since the total energy $E_{\rm{plasma}}$ scales linearly with the plasma's density. The density function for the ambient matter field may thus be modelled similar to \citet{blandford1979relativistic_conicalJet} via
\begin{equation}
	n_{\rm{plasma}}(r) = \frac{\kappa_{\rm{plasma}}}{r^2}
\end{equation}
using a normalization constant $\kappa_{\rm{plasma}}$.

\subsection{Photon Fields}
\label{subsec:photonFields}

\subsubsection{Accretion Disc Photons}
\label{subsubsec:accDiscPhotons}

AGN are believed to be powered by the accretion of matter onto a supermassive black hole (\citet{mckinney2006general_blackHolesAccretionJet}). Such an accretion disc contains hot matter which radiates thermal photons into the AGN's surroundings. The spectral energy distribution of thin accretion discs is described by a black body spectrum of temperature $T_0$ given in the rest frame of the AGN as pointed out in the work of \citet{li2005blackbodyAccDisc}. The black body radiation field will be Doppler-de-boosted in the plasmoid frame as it is receding from the accretion disc with a Lorentz factor $\Gamma$. The Doppler(-de)-boost from the accretion disc system at rest into the plasmoid rest frame is carried out according to \citet[Eq.\ 11]{lee2017relativisticBlackbody}
\begin{equation}
	n_{\gamma, \tiny\mbox{black body}}^{(\rm{plasmoid})}(\epsilon_\gamma,\delta) = \frac{8\pi }{\left(hc\right)^3}\cdot\frac{(\delta\epsilon_\gamma)^3}{\exp\{\frac{\epsilon_\gamma}{k_bT_0\delta}\} - 1}
	\label{eq:photonDensityBlackbodyDeBoost}
\end{equation}
in which $\delta$ is the Doppler factor
\begin{equation}
    \delta(\Gamma)=\left[\Gamma-\sqrt{\Gamma^2-1}\cos{\theta}\right]^{-1}
    \label{eq:dopplerFunction}
\end{equation}
with $\theta$ as the angle of the system's direction of motion and the observer's line of sight. Hence, the temperature of a black body spectrum gets modified by a Doppler factor and can, as it is the case for a receding frame of reference, appear decreased when transformed into the plasmoid rest frame.

The photon density will decrease while the plasmoid propagates along the jet axis due to the increase in distance. Hence, the photon field itself is modelled by a separation ansatz
\begin{equation}
	n_\gamma(\epsilon_\gamma, r) = n_{\gamma,0}(\epsilon_\gamma) \cdot \mathcal{S}(r)
	\label{eq:sepAnsatz}
\end{equation}
in which the factor $n_{\gamma,0}(\epsilon_\gamma)$ describes the spectral scaling of the photon number density and the scaling function $\mathcal{S}(r)$ the field's geometric scaling in dependence of the location of the plasmoid inside the jet. This scaling function fulfills the requirement that it be unity where it assumes its highest value in order to represent an up- or down-scaling of the energy-dependent photon number density. For simplicity, the accretion disc is modelled to emit photons homogeneously over its entire surface area $A_{\rm{acc}}=\pi\left( R_{\rm{acc}}^2-(3R_s)^2\right)$ in which $R_{\rm{acc}}$ is the outer radius of the accretion disc and $3R_s$ the innermost stable circular orbit (ISCO) for non-massless particles around a black hole. The outer radius of the accretion disc satisfies that $R_{\rm{acc}}\gg 3R_s$, yet enters as a rather free parameter whose value and impact on the evolution of the plasmoid will be reviewed in the subsequent sections.

\begin{figure*}
	\includegraphics[width=\textwidth]{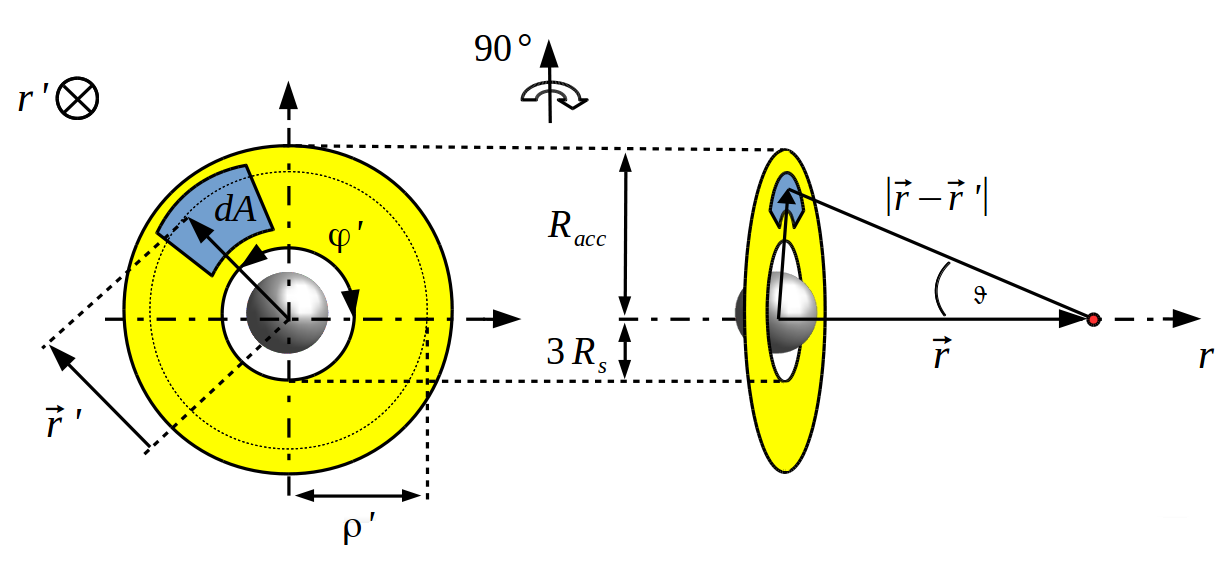}
	\caption{Sketch of a thin accretion disc with minimal radius $3R_s$ and maximal radius $R_{\rm{acc}}$ around a black hole. The left panel shows the accretion disc face-on with $\rho\,'$ and $\varphi\,'$ as polar coordinates in the plane of the accretion disc spanning up a vector $\vec{r}'$ of length $\rho\,'$ pointing towards an infinitesimal surface element $\mathrm{d}A$. The right panel shows an isometric edge-on view of the same accretion disc with the $r$-axis pointing along the jet. A vector $\vec{r}$ points from the origin towards the point at which the photon field intensity is to be evaluated. This point is separated by a distance of $|\vec{r}-\vec{r}'|$ from $\mathrm{d}A$ under an angle $\vartheta$ towards the surface of the accretion disc.}
    \label{fig:figure2_discScalingSketch}
\end{figure*}

The photon intensity $I(r)$ that is emitted by such an accretion disc can be estimated by superposing infinitesimally small regions of the radiating volume with power density $\varrho_P$ inside a volume $\mathrm{d}V$. If $\vec{r}$ is the position in space at which the intensity should be evaluated and $\vec{r}'$ the vector pointing towards the photon-emitting region, the intensity can be found by 
\begin{equation}
	I(r) = \frac{1}{4\pi}\int\frac{\varrho_P(\vec{r}')}{|\vec{r}-\vec{r}'|^2}\,\mathrm{d}V\left(\vec{r}'\right)
	\label{eq:intensityIntegral}
\end{equation}
as illustrated in Fig.~\ref{fig:figure2_discScalingSketch}.
If the total radiation is emitted with total power $P_0$ equally over the accretion disc surface, the power density of the accretion disc within an infinitesimal volume element $\mathrm{d}V$ is given by
\begin{equation}
    \begin{split}
	    \varrho_P\left(\rho\,', \varphi\,', r\,'\right)\mathrm{d}V
	    &=\frac{P_0}{A_{\rm{acc}}}\cdot H(\rho\,'-3R_s)\quad\times\\
	    \times\quad H(&R_{\rm{acc}}-\rho\,')\cdot\delta(r\,'-0)\,\rho\,'\,\mathrm{d}\rho\,'\,\mathrm{d}\varphi\,'\,\mathrm{d}r\,'
    \end{split}
\end{equation}
where $\rho\,'$ and $\varphi\,'$ are the polar coordinates spanning up the surface of the accretion disc and $r'=0$ the position of the accretion disc on the jet axis. Note that primed variables indicate coordinates pointing towards $\mathrm{d}A$ and that $r$ and $r'$ correspond to the cylindrical coordinate $z$ which is not used in this context in order to prevent confusion with the name convention for redshift. From geometry directly follows that $|\vec{r}-\vec{r}'|^2=\rho\,'^{\,2} + (r-r')^2$ and hence follows for the intensity of the entire accretion disc
\begin{equation}
    \begin{split}
	    I(r)&=\frac{P_0}{4\pi A_{\rm{acc}}}\int_{-\infty}^{\infty}\mathrm{d}r\,'\int_{0}^{2\pi}\mathrm{d}\varphi\,'\int_{3R_s}^{R_{\tiny\mbox{acc}}}\,\mathrm{d}\rho\,'\frac{\rho\,'\,\delta(r\,')}{\rho\,'^{\,2}+(r-r')^2}\\
	    &=\frac{P_0}{4A_{\rm{acc}}}\ln\left(\frac{R_{\rm{acc}}^2+r^2}{\left(3R_s\right)^2+r^2}\right)=\frac{1}{4}\sigma_{\rm{SB}}T_0^4\ln\left(\frac{R_{\rm{acc}}^2+r^2}{\left(3R_s\right)^2+r^2}\right)~~.
    \end{split}
    \label{eq:accDiscIntensity}
\end{equation}
where in the last step we have used the expression for the radiated power of a black body with $\sigma_{\rm{SB}}$ being the Stefan-Boltzmann constant. One may impose the same geometrical scaling on a scaling function $\mathcal{S}_{\rm{disc}}(r)$ of the photon density as an infinitesimally small photon-emiting region of the accretion disc is equally affected by a $r^{-2}$-loss in intensity. Using Eq.~(\ref{eq:accDiscIntensity}) the scaling function can be modelled via
\begin{equation}
    \mathcal{S}_{\rm{disc}}(r)=\kappa_{\mathcal{S},\rm{disc}}\cdot\ln\left(\frac{R_{\rm{acc}}^2+r^2}{\left(3R_s\right)^2+r^2}\right)
	\label{eq:scalFunctAccDisc}
\end{equation}
where $\kappa_{\mathcal{S},\rm{disc}}$ is a normalisation constant. Due to geometry one finds that the maximum of the scaling function has to exist at $r=0$ which implies
\begin{equation}
	\mathcal{S}_{\rm{disc}}(r=0)=1 ~~\Rightarrow~~ \kappa_\mathcal{S,\rm{disc}}=\left[2\ln\left(\frac{R_{\rm{acc}}}{3R_s}\right)\right]^{-1}~.
\end{equation}
To understand the behaviour of this scaling function one may investigate it around regions near and far from the accretion disc. The former case is given where $r\ll R_{\rm{acc}}$ in which case one may Taylor-expand Eq.~(\ref{eq:scalFunctAccDisc}) around small $r\ll R_{\rm{acc}}$ yielding in first order $\mathcal{S}_{\rm{disc}}(r\ll R_{\rm{acc}})\propto 1 - \kappa_\mathcal{S,\rm{disc}}(r/3R_s)^2$ which is approximately unity for sufficiently small $r$. For distances far from the accretion disc the scaling function may be expanded around small values $R_{\rm{acc}}\ll r$ which reproduces the scaling of a point-like source with $\mathcal{S}_{\rm{disc}}(R_{\rm{acc}}\ll r)\propto (R_{\rm{acc}}/r)^2$.

\subsubsection{Synchrotron Emission}
\label{subsubsec:SynchField}

The population of relativistic electrons is deflected by the ambient turbulent magnetic field and generates an isotropic photon field of synchrotron radiation in the rest frame of the emitting plasmoid. The electrons' number density is modelled to be energetically power-law distributed according to
\begin{equation}
	\frac{\mathrm{d}n_e}{\mathrm{d}E_e} = \kappa_e\cdot\left(\frac{E_e}{m_ec^2}\right)^{-\alpha_e}
\end{equation}
with $\kappa_e$ as a normalisation constant in units of $\rm{m}^{-3}~\rm{J}^{-1}$ and $m_e$ the electron rest mass. This spectrum is normalised via the electron energy density $U_e$ which in return can be rewritten using Eq.~(\ref{eq:energyBudget})
\begin{equation}
    U_e = \frac{U_b}{\chi+1} = \kappa_e m_e^2c^4 \int_{\gamma_{e,\min}}^{\gamma_{e,\max}} \gamma_e^{1 - \alpha_e} \mathrm{d}\gamma_e
    \label{eq:calcElectronEnergyDensity}
\end{equation}
with $\gamma_{e}$, $\gamma_{e,\min}$, $\gamma_{e,\max}$ as the (minimal / maximal) Lorentz factor of the electron population whose ranges of values will be discussed in Sec.~\ref{subsec:parameterSetup}. The well-known solution to the frequency-dependent synchrotron emissivity in the interval $\nu+\mathrm{d}\nu$ is in return a power-law as described in \citet{longair2011}. This gives
\begin{equation}
	j_s(\nu) = \frac{\sigma_Tc\kappa_e}{3\mu_0}m_ec^2\beta_e^2\cdot\left(\frac{e}{2\pi m_e}\right)^{\frac{\alpha_e-3}{2}}B^{\frac{\alpha_e+1}{2}}\cdot\nu^{\frac{1-\alpha_e}{2}}
	\label{eq:synchEmissivity}
\end{equation}
in units of $\rm{J}~\rm{s}^{-1}~\rm{m}^{-3}~\rm{Hz}^{-1}$, $\sigma_T$ the Thomson cross section, $\beta_e^2=1-\gamma_e^{-2}\approx 1$, and $\kappa_e$ being directly calculable from Eq.~(\ref{eq:calcElectronEnergyDensity}). As emissivities are a measure of energy per unit frequency, unit time, and unit volume, an estimate of the number density of synchrotron photons injected per unit time $\Dot{n}_{\gamma,\rm{synch}}$ can be obtained by
\begin{equation}
    \Dot{n}_{\gamma,\rm{synch}}(\epsilon_\gamma) =  j_s(\nu=\epsilon_\gamma/h)\cdot\nu/\epsilon_\gamma\cdot \equiv j_s(\epsilon_\gamma)/h
\end{equation} 
in units of $\rm{m}^{-3}\rm{s}^{-1}$ and with $\epsilon_\gamma$ as the energy of the synchrotron photon. To achieve the absolute number density of synchrotron photons we integrate the expression above over a time characteristic for the emission of synchrotron radiation. On average, electrons would have radiated most of their energy into synchrotron radiation after a time of
\begin{equation}
    t_{\rm{synch}} = \frac{3m_ec^2}{4\sigma_T c U_b \left<\gamma_e\right>^2}
\end{equation}
in which $\left<\gamma_e\right>$ is the average Lorentz factor of the electron population that is emitting synchrotron radiation. After simplifying all terms, the number density of photons of the synchrotron field reads as
\begin{equation}
    \begin{split}
        {n}_{\gamma,\rm{synch}}(\epsilon_\gamma)=&\frac{\beta_e^2}{4\mu_0h(\chi + 1)\left<\gamma_e\right>^2}\cdot\left(\frac{e}{2\pi m_e}\right)^{\frac{\alpha_e-3}{2}}\cdot\quad\times\\  \times\quad&\left[\int_{\gamma_{e,\min}}^{\gamma_{e,\max}} \gamma_e^{1 - \alpha_e} \mathrm{d}\gamma_e\right]^{-1}\cdot B^{\frac{1+\alpha_e}{2}}\cdot\left(\frac{\epsilon_\gamma}{h}\right)^{\frac{1-\alpha_e}{2}}
    \end{split}
    \label{eq:numberDensitySynchrotron}
\end{equation}
in units of $\rm{m}^{-3}$. Note that the synchrotron photon number density is proportional to the synchrotron emissivity itself, thus also resembling a power-law.

Due to the dependence of the synchrotron photon number density on the magnetic field through which the emitting electron population is propagating, the synchrotron field itself will vary accordingly. Given a dependence on the magnetic field as described in Eq.~(\ref{eq:synchEmissivity}) and a scaling of the magnetic field along the jet axis derived in Eq.~(\ref{eq:BfieldScaling}), one acquires the scaling of the synchrotron photon number density over distance inside the jet proportional to
\begin{equation}
	n_{\gamma,\rm{synch}}\propto B^{\frac{1+\alpha_e}{2}} \mbox{ ,}~~ B(r)\propto\frac{1}{r} ~\Rightarrow~ n_{\gamma,\rm{synch}}(r)\propto r^{-\left(\frac{1+\alpha_e}{2}\right)}~.
	\label{eq:scalFuncSynchField}
\end{equation}
With a separation ansatz of the spectral and geometrical scaling as in Eq.~(\ref{eq:sepAnsatz}) one can define the scaling function of the synchrotron field along the jet axis via
\begin{equation}
    \mathcal{S}_{\rm{synch}}(r)=\kappa_{\mathcal{S},\rm{synch}}\cdot r^{-\left(\frac{1+\alpha_e}{2}\right)}
	\label{eq:scalFunctSynchField}
\end{equation}
which is normalised by defining the maximum value of the scaling function to be unity at $r_{\rm{start}}$
\begin{equation}
	\mathcal{S}_{\rm{synch}}(r=r_{\rm{start}})=1 ~~~\Rightarrow~~~ \kappa_{\mathcal{S},\rm{synch}}=r_{\rm{start}}^{\frac{1+\alpha_e}{2}}~~.
\end{equation}

\subsection{Simulation Setup}
\label{subsec:simSetup}

The model presented in Sec.~\ref{sec:model} has been simulated in 3+1 dimensions by modifying the publicly available {\small CRPROPA} propagation framework (version 3.1+) by \citet{batista2016crpropaMain}. This enables us to consider photon fields as e.g.\ described in Eqs.~(\ref{eq:photonDensityBlackbodyDeBoost}) and (\ref{eq:synchEmissivity}) as well as dependence of these fields on space and time as described in Eqs.~(\ref{eq:scalFunctAccDisc}) and (\ref{eq:scalFuncSynchField}). The environmental properties and interactions that have been taken into account for the simulation of the plasmoid are:

\begin{itemize}
    \item Photo-pion production of high-energy protons or neutrons with photon backgrounds represented by the accretion disc photon field and the synchrotron photon field produced by the ambient electron population:
    \begin{equation}
        p/n_{\rm{ high-energy}} + \gamma_{\rm{acc/synch}} \rightarrow \pi^{\pm/0} + X\,.
        \label{eq:processPGamma}
    \end{equation}
    Photo-hadronic interactions are simulated using the corresponding {\small CRPROPA} module which in return makes use of the SOPHIA code written by \citet{SOPHIA_original}. SOPHIA takes into account the nine most relevant nuclear resonances and their decay channels of each the proton and the neutron whilst also considering contributions from secondary $\rho$, $\omega$, and $\eta$ mesons. The most common interaction within this channel is the transformation of a proton to a neutron under emission of pions decaying into neutrinos and gamma-rays as well as the production of $p\bar{p}$ and/or $n\bar{n}$ pairs.
    \item Proton-proton interactions of high-energy primary protons and protons of the ambient plasma according to
    \begin{equation}
        p_{\rm{high-energy}} + p_{\rm{ambient}} \rightarrow \pi^{\pm/0} + X\,.
        \label{eq:processPP}
    \end{equation}
    are taken into account. In analogy to photo-hadronic interactions, pions contribute to the resulting secondary particle populations of neutrinos and gamma-rays.
    \item Electron-pair production by absorption of a high-energy gamma-ray with a background photon being either from the accretion disc or electron synchrotron photon field as described by
    \begin{equation}
        \gamma_{\rm{high-energy}}+ \gamma_{\rm{acc/synch}}\rightarrow e^{+} + e^{-}
        \label{eq:processGammaGamma}
    \end{equation}
    \item Proton synchrotron radiation
    \item Decay of unstable particles
\end{itemize}

\subsection{Parameter Setup}
\label{subsec:parameterSetup}

The radial extent of the plasmoid was chosen to be $R=10^{13}~\rm{m}$ in compliance with the work of  \citet{eichmann2010plasmoidLeptonic, eichmann2011duration, eichmann2012plasmoidHadronic} and comparable to \citet{giannios2009fast_blobSize_blobProcesses} into which the primary population of protons has been injected isotropically, homogeneously, and instantaneously with each proton having an initial energy of $E_{p,\rm{inj}}=10^8~\rm{GeV}$, entering as a free parameter. The plasmoid starts propagating at a distance of $r_{\rm{start}}=10^{14}~\rm{m}~=10~R$ from the black hole centre with a Lorentz factor of $\Gamma=10$ (see e.g.\ \citet{asada2013discovery_VLBIgamma10, giannios2013reconnection_PlasmoidGamma} which remains constant in the entire simulation. The simulation terminates after the plasmoid has propagated for a distance of $10~\rm{pc}$ in its own reference system which corresponds to a distance from the black hole of $r_{\rm{end}}=r_{\rm{start}}+10~\rm{pc}=3.1\cdot10^{17}~\rm{m}$ in the plasmoid system or $r_{\rm{end}}/\Gamma$ in the rest frame of the black hole. The simulation evolved with a spatial resolution of $\Delta r=10^{-2}R$ which, for particles moving approximately at the speed of light, corresponds to a temporal resolution of $\Delta t = \Delta r/c\approx3.3\cdot 10^{2}$ seconds. The environmental parameters of the plasmoid are as follows:
The turbulent magnetic field has a Kolmogorov type wave spectrum as pointed out by \citet{biermann1987synchrotron_BneedstobeKolmogorov} and was designed to achieve a correlation length of $L_c=\Delta r=10^{-2}R$ such that the model remains in the isotropic regime in each step of the simulation. The root-mean-square value of the turbulent magnetic field was chosen to be $B=10^{-4}~\rm{T}=1~\rm{G}$ as e.g.\ used in \citet{inoue1996electron_B1g, eichmann2012plasmoidHadronic, christie2018radiative_EquilibriumPlasmoids_B1G} which governs the absolute amount of energy available to the energetically significant constituents of the jet as described in Eq.~(\ref{eq:energyBudget}). The corresponding magnetic energy density thus is in the order of $U_b\approx 4\cdot10^{-3}~\rm{J/m}^3$ and we assume a lepton-dominated jet such that $\chi\ll 1$ which is modelled to be $\chi=1/100$. The magnetic field strength scales proportional to $r^{-1}$ as pointed out in Sec.~\ref{subsec:magneticField} to reflect energy conservation along the jet. The black body photon field of the accretion disc is modelled to possess a single temperature in order to keep the model simple with a temperature of $T_{\rm{disc}}=10~\rm{eV}/k_b$ which lies in the range of typical temperatures as pointed out in \citet{ross1992AccDiscSpectra, di1998Tbb, eichmann2012plasmoidHadronic}. The field gets Doppler-de-boosted into the plasmoid rest frame as the plasmoid is receding from the accretion disc at a relativistic velocity. In the case of the observer being in a system receding from a source at rest, we approximate the angle in the Doppler function in Eq.~(\ref{eq:dopplerFunction}) as $\vartheta=\pi$, neglecting further angular dependence as shown in \citet{potter2012synchrotron_angularDependenceDoppler} and in particular at distances $r>R_{\rm{acc}}$ (comp. Eq.~(\ref{eq:scalFunctAccDisc})) we assume light to reach the plasmoid approximately parallel. The angle-averaged Doppler factor thus becomes $\delta\approx 1/10$ for an observer at small angles receding at $\Gamma=10$ which, hence $\delta\approx 1/\Gamma$ and according to the photon density of a relativistically modified black body spectrum in Eq.~(\ref{eq:photonDensityBlackbodyDeBoost}), lets the temperature of the accretion disc appear cooler by about one order of magnitude with a resulting average photon energy of 7.6 eV in the plasmoid frame. The geometric scaling of the accretion disc field in Eq.~(\ref{eq:scalFunctAccDisc}) invokes the inner and outer radii of the accretion disc for which we used $3R_s$ and $R_{\rm{acc}}=10^3R_s\approx 10\,R$ with the Schwarzschild radius $R_s=2GM/c^2\approx 3\cdot10^{11}~\rm{m}$ of a $M=10^8 M_{\rm{sol}}$ solar mass black hole as pointed out in \citet{starling2004constraints_Racc}. The second photon field in this model is represented by the synchrotron photon field generated by an electron population moving with the plasmoid. The individual trajectories of these electrons have not been tracked in the simulation yet the photon field they generate is taken into account. The parameters of the electron population used to normalise the electron energy density in Eq.~(\ref{eq:calcElectronEnergyDensity}) are $\alpha_e=-2.6$ as used in \citet{cavagnolo2010relationship_spectralIndexElectrons2.6} for the spectral index and $\gamma_{e,\min}=10$ as well as $\gamma_{e,\max}=10^6$ for the minimal and maximal Lorentz factor. Values for $\gamma_{e,\min/\max}$ range from values from 1 to 100 and $10^3$ to $10^7$ in literature as e.g.\ in \citet{petropoulou2016blazar_ymin1to100_ymax1e4, ahnen2018extreme_electron_ymin100_ymax1e7} yet have a comparably small impact on the number density due to the usually steep shape of electron spectra. The chosen spectral index results in a geometric scaling along the jet axis proportional to $r^{-1.8}$. Finally, the ambient proton plasma was modelled with a density of $n_{0, \rm{plasma}}=10^{15}~\rm{m}^{-3}$. We recognise such densities to be very high compared to what arguments of energetic equipartition suggest in our model yet have chosen this value to achieve a scenario of a maximal proton density in which, on average, a fraction of $1/100$ of energy lost by primaries into secondary particles goes into proton-proton interactions. By this approach, losses by the $pp$-channel do not significantly influence the energetic evolution of the plasmoid's primary particles such that one would have achieved nearly identical results for their propagation if proton-proton interactions would have been neglected at all. The results of changing the parameter $n_{0,\rm{plasma}}$ are investigated in Sec.~\ref{subsubsec:densityOfProtonPlasma}. Furthermore, the proton plasma is subject to a decrease in density proportional to $r^{-2}$ due to geometric effects as described in Sec.~\ref{subsec:ambientMatter} rendering an efficient production of secondary particles from $pp$-interactions in our model the more unlikely, the larger the distance of the plasmoid from the central black hole.

\begin{table}
	\centering
	\caption{Parameters used in the simulation. All parameters are given in the plasmoid rest frame with the exception of the accretion disc parameters.}
	\label{tab:example_table}
	\begin{tabular}{lcc} 
		\hline
		Parameter & Symbol & Value\\
		\hline
		Plasmoid Radius & $R$ & $10^{13}~\rm{m}$\\
		Plasmoid Propagation Start & $r_{\rm{start}}$ & $10^{14}~\rm{m}$\\
		Plasmoid Propagation End & $r_{\rm{end}}$ & $r_{\rm{start}}+10~\rm{pc}$\\
		Plasmoid Lorentz Factor & $\Gamma$ & $10$\\
		Magnetic Field Initial RMS Value & $B_0$ & $1~\rm{G}$\\
		Proton (primary) Initial Energy & $E_{p,\rm{inj}}$ & $10^8~\rm{GeV}$\\
		Proton Target Density (up-scaled) & $n_{0,\rm{plasma}}$ & $10^{15}~\rm{m}^{-3}$\\
		Electron Minimal Lorentz Factor & $\gamma_{e,\min}$ & $10$\\
		Electron Maximal Lorentz Factor & $\gamma_{e,\max}$ & $10^6$\\
		Electron Spectral Index & $\alpha_e$ & $2.6$\\
		Energy Density Ratio $U_p/U_e$ & $\chi$ & $1/100$\\
		Accretion Disc Inner Radius & $3R_s$ & $8.86\cdot 10^{11}~\rm{m}$\\
		Accretion Disc Outer Radius & $R_{\rm{acc}}$ & $10^{14}~\rm{m}$\\
		Accretion Disc Temperature & $T_{0}$ & $10$ eV/k$_b$\\
		\hline
	\end{tabular}
	\label{tab:parameters}
\end{table}

We could increase the computational simulation efficiency by not propagating the plasmoid through the entire distance from $r_{\rm{start}}$ to $r_{\rm{end}}$. Instead, the plasmoid was modelled at rest with the environment being Lorentz-transformed and propagated through the plasmoid. For the parameters used in this work this allowed for the used grid points to be reduced from $(\pi R^2 \cdot (r_{\rm{end}}-r_{\rm{start}} - 2R) + 4/3\pi R^3) \cdot 1/(\Delta r)^3 \approx 9.7\cdot 10^{10}$ to $4/3\pi R^3\cdot 1/(\Delta r)^3 \approx 4.2\cdot 10^6$.

\section{Results}
\label{sec:results}

\subsection{Relative Importance of Processes}
\label{subsec:importanceOfProcesses}

The parameters mentioned above result in certain processes being dominant at specific regimes of energy, time, and density. To evaluate the importance of these processes during the evolution of the plasmoid, we infer the average number of interactions $\tau$ of a particle with given energy propagating a characteristic length scale through the plasmoid which we chose to be one plasmoid radius. This concept is analog to that of the optical depth, hence we adopt the symbol $\tau$ as well. The average number of interactions of a particle per propagation length of a process can be acquired by multiplying its energy-dependent interaction rate with a length of propagation. Noting that the interaction rate is equivalent to the inverse of the mean free path $\lambda$, we have calculated $\tau$ as follows:
\begin{enumerate}
    \item For photo-hadronic interactions,
    \begin{equation}
        \tau(E_p)=\frac{R}{\lambda\left(E_p\right)}=R\int_{\epsilon_{\gamma,\min}}^{\epsilon_{\gamma,\max}} \frac{\rm{d}n_\gamma(\epsilon_\gamma)}{\rm{d}\epsilon_\gamma}\cdot\bar{\sigma}\left(\frac{2E_p\epsilon_\gamma}{m_pc^2}\right)\rm{d}\epsilon_\gamma
        \label{eq:tauPGamma}
    \end{equation}
    with the photon field's number density per energy interval $\rm{d}n_\gamma/\rm{d}\epsilon_\gamma$ and the energy-weighted average cross section
    \begin{equation}
        \bar{\sigma}(\xi)=\frac{2}{\xi^2}\int_0^{\xi}\sigma_{p\gamma}(\epsilon_\gamma)\cdot\epsilon_\gamma~\rm{d}\epsilon_\gamma
    \end{equation}
    as described in the work of \citet[Eqs.~(1), (2)]{crpropaMFP} whose approach is used to calculate photo-hadronic mean free paths for the {\small CRPROPA} framework. Note that the proton-photon cross section $\sigma_{p\gamma}$ has been achieved using the SOPHIA code (for an overview see \citet[Fig.~1]{kelner2008protonGamma}).

    \item In the case of proton-proton interactions we have calculated $\tau$ by multiplying the number density of the proton plasma $n_{0,\rm{plasma}}$ by the inelastic cross section for proton-proton interactions and the radius of the plasmoid. The cross section is taken from \citet[Eq.~(79]{kelner2006protonProton} who calculated a parameterisation of
    \begin{equation}
        \sigma_{pp}(E_p)=\left(34.3 + 1.88 L + 0.25 L^2\right)\left[1-\left(\frac{E_{\rm{th}}}{E_p}\right)^4\right]^2
        \label{eq:sigmaPP}
    \end{equation}
    given in mb with $L=\ln(E_p/\rm{TeV})$ and $E_{\rm{th}}=1.22~\rm{GeV}$ as the threshold energy for inelastic proton-proton collisions.

    \item For $\gamma\gamma$-absorption we use the approach of \citet{gould1967pair} who calculate the average number of interactions in a similar approach as in Eq.~(\ref{eq:tauPGamma}) (comp.\ Eq.~(8 of their work), yet using the cross section for photon-photon collisions of a photon with energy $E_\gamma$ with a background photon of energy $\epsilon_\gamma$
    \begin{equation}
        \sigma_{\gamma\gamma}(E_\gamma,\epsilon_\gamma)=\frac{3\sigma_T}{16}\left(1-\eta^2\right)\left[ \left(3-\eta^4\right)\ln\left(\frac{1+\eta}{1-\eta}\right)-2\eta\left(2-\eta^2\right)\right]
        \label{eq:sigmaGammaGamma}
    \end{equation}
    with $\eta(E_\gamma,\epsilon_\gamma)=[1-4m_e^2c^4/(E_\gamma\epsilon_\gamma)]\,^{1/2}$. Ready-to-use Python implementations to calculate the interaction rates of the above mentioned two processes currently exist in the official {\small CRPROPA} side-repository\footnote{\url{https://github.com/CRPropa/CRPropa3-data}}.
\end{enumerate}

\begin{figure}
	\includegraphics[width=0.48\textwidth]{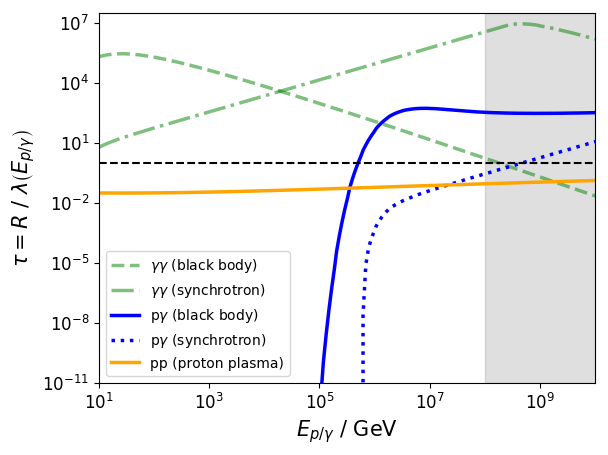}
	\caption{Average number of interactions / optical depth $\tau$ (y-axis) at $r_{\rm{start}}$ of an incident test proton (photon) of energy $E_{p\gamma}$ (x-axis) propagating a length of $1R=10^{13}~\rm{m}$ through an environment with properties as specified in the model section. The curves show photo-hadronic interactions ($p\gamma$), proton-proton interactions ($pp$) and the creation of electron-positron pairs from the interaction of a high-energy gamma-ray and one photon of the two ambient photon backgrounds ($\gamma\gamma$). Energies of particles above $10^8~\rm{GeV}$ are shaded as these energies have not been achieved by any particle in the simulation. For particles with energies above the threshold of photo-pion production at roughly $10^5~\rm{GeV}$ energetic losses into $p\gamma$-interactions become dominant leading to mostly photo-hadronic particle emission.}
	\label{fig:figure3_opticalDepth}
\end{figure}

The energy-dependent, average numbers of interactions of all processes considered in this work are illustrated in Fig.~\ref{fig:figure3_opticalDepth} in which the x-axis denotes the energy of a primary proton or gamma-ray of energy $E_{p/\gamma}$. Energies of particles above $10^8~\rm{GeV}$ are shaded as these energies have not been achieved by any particle in the simulation. All curves are given at $r=r_{\rm{start}}$ and need to be scaled according to the scaling functions corresponding to the respective process if to be evaluated at positions more distant from the central black hole.
\begin{itemize}
    \item The blue, solid line represents the number of interactions of photo-hadronic pion production (Eq.~(\ref{eq:processPGamma})) as specified in Eq.~(\ref{eq:tauPGamma}) with the accretion disc black body photon field, whereas the blue, dotted line corresponds to photo-hadronic pion production with the synchrotron photon field emitted by the ambient, relativistic electron population. Both interactions become possible at energies approximately above $E_p>10^{5}~\rm{GeV}$ yet interactions with the black body field are by several orders of magnitude more often encountered compared to interactions with the synchrotron field. Note that photo-hadronic pion production is not only possible in protons but also in neutrons, yet their cross sections with photons are nearly identical at the energies considered in our simulation.
    \item The orange, solid line illustrates the average number of interactions for proton-proton interactions (Eq.~(\ref{eq:processPP})) using the cross section in Eq.~(\ref{eq:sigmaPP}) which is approximately constant across all relevant energies yet linearly depends on the number density $n_{\rm{gas}}$ of the ambient proton plasma. Hence, proton-proton-interactions are the dominant interaction process for protons below the energy threshold for photo-hadronic interactions in this model and pose a permanent source of secondary particles in the considered energy regime.
    \item Gamma-rays, which in our work all appear as secondary particles from the interactions considered, can be absorbed upon interaction with photons from the two background photon fields causing the creation of an electron-positron pair (Eq.~(\ref{eq:processGammaGamma})). The optical depth of this process with respect to the black body photon field is described by the green, dashed curve using the cross section specified in Eq.~(\ref{eq:sigmaGammaGamma}) which renders the plasmoid optically thick to gamma-rays below $\approx 10^8~\rm{GeV}$ over a distance of $R$ and optically thin to gamma-ray energies above this value. Conversely, the plasmoid is optically thick to gamma-rays above $\approx 10~\rm{GeV}$ due to interactions with the synchrotron photon field whereas this process is less dominant below this energy. Both photon fields form a ``sweet spot" around a gamma-ray energy of $\approx 10~\rm{TeV}$ at which the combined optical depths of both processes is the lowest. Hence, in this model, gamma-rays of this energy are the most likely to escape the plasmoid as opposed to a larger number of photons undergoing $e^{+}/e^{-}$ pair production within the plasmoid.
\end{itemize}




\subsection{Evolution of the Plasmoid}
\label{subsec:plasmoidEvolution}

\begin{figure*}
	\includegraphics[width=\textwidth]{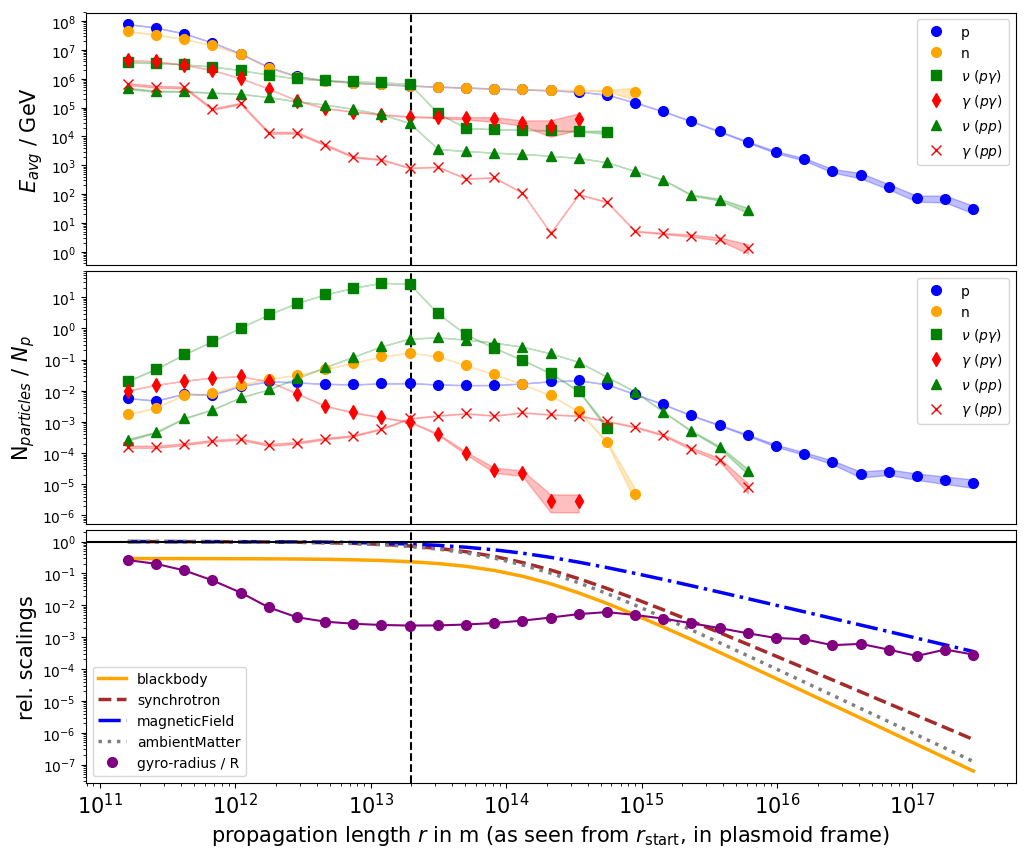}
	\caption{Evolution of the properties of the relativistic particle population in a plasmoid of radius $R=10^{13}~\rm{m}$ during its propagation along the jet axis from $r_{\rm{start}}=10^{14}~\rm{m}$ to $r_{\rm{end}}=3\cdot 10^{17}~\rm{m}=10~\rm{pc}$ with a Lorentz factor of $\Gamma=10$. Values on the x-axis are measured from $r_{\rm{start}}$ and show the distance (or time of divided by $c$) for which the plasmoid has propagated in its own rest frame. The upper panel shows the average energy of various particles in GeV as seen by an observer covering the entire surface of the plasmoid, whereas the panel in the middle displays the number of particles detected in relation to the number primary particles (all protons) injected. Particles produced in photo-hadronic interactions are labels $p\gamma$ as opposed to particles produced in proton-proton-interactions with label $pp$. The lower panel illustrates the scaling functions of the accretion disc field, the electron synchrotron photon field, the magnetic field, the proton plasma density as well as the average gyro-radius of detected protons in multiples of the plasmoid radius.}
    \label{fig:figure4_plasmoidEvolution}
\end{figure*}

In order to make comparisons to the scales of the plasmoid easier, we will refer to distances along the jet in multiples of the plasmoid radius $R=10^{13}~\rm{m}$ as well as to times in multiples of $R/c=3.33\cdot10^4~\rm{s}\approx 9.25~\rm{hours}$. The main results of this work are summarised in Fig.~\ref{fig:figure4_plasmoidEvolution}, covering the entire propagation from $r_{\rm{start}}$ to $r_{\rm{end}}$. Note that the x-axis starts measuring from $r_{\rm{start}}$ on in order to illustrate effects occurring short after the start of propagation and is also given in the plasmoid rest frame. Hence, in order to achieve the absolute distance $r_{\rm{jet}}$ in the jet from the central black hole, one has to take the starting point as well as the plasmoid's Lorentz factor into account $r_{\rm{jet}}=(r_{\rm{start}}+r)/\Gamma$. All further distances and times are given in the plasmoid rest frame and need to be transformed accordingly using the equation above. 
The figure itself consists of three panels each covering a particular aspect of the plasmoid's evolution. The upper panel shows the average energy in GeV of several species of particles as measured when leaving the plasmoid, i.e.\ as seen by an observer sitting on and covering the entire surface of the plasmoid. The panel in the middle counts the absolute numbers of these particles relative to the total number of primary protons that have been injected into the plasmoid volume as primary particles. Finally, the lower panel reflects the relative scaling of all target fields which are described in the scaling of the black body accretion disc field in Eq.~(\ref{eq:scalFunctAccDisc}), the synchrotron field in Eq.~(\ref{eq:scalFuncSynchField}), the magnetic field in Eq.~(\ref{eq:BfieldScaling}), and the ambient proton plasma in Eq.~(\ref{eq:ambientMatterFieldScaling}). A fifth curve describes the average gyro-radius $R_g(E_p, r)\approx E_p/(ecB(r))$ of all protons remaining inside the plasmoid in multiples of $R$.\\

\subsubsection{Protons} 
\label{subsubsec:protons}

Upon injection into the plasmoid, protons (blue circles) dominantly start undergoing interactions with the black body field as the number of average interactions of this process is roughly in the order of 100 at energies above $10^6$ GeV. Due to energy going into the production of secondary particles, the average energy of the proton population decreases by two orders of magnitude from $10^8~\rm{GeV}$ to roughly $10^6~\rm{GeV}$ within the first $\approx 0.5R/c$ seconds (comp.\ Fig.~\ref{fig:figure7_plasmoidImages} upper vs.~ middle panel) while producing neutrons as also pointed out in \citet{atoyan2003neutral_n1e1_neutralCurrents}. Beyond this energy, the probability of interactions of protons with accretion disc photons rapidly decreases until an energetic plateau is reached near the threshold for photo-pion production. This plateau is reached well before geometric effects become dominant as at energies between $E_p \approx 10^6~\rm{GeV}$ to $5\cdot10^5~\rm{GeV}$ there is a decrease of two orders of magnitude in the average amount of interactions of $p\gamma$-interactions with the accretion disc field (comp.\ Fig.~\ref{fig:figure3_opticalDepth}). In comparison, photo-hadronic interactions with the synchrotron field are generally by two up to five orders of magnitude suppressed such that most protons lose their energy into interactions with the black body field. Near the threshold energy for $p\gamma$-interactions with the synchrotron field at roughly $10^6~\rm{GeV}$, a proton/neutron in our model would still interact by a factor of 100 more likely with the accretion disc field, rendering the plasmoid quasi-transparent for nucleons for photo-pion production with synchrotron radiation at any given time of the simulation.

For energies below the threshold for photo-hadronic interactions with the black body field at roughly $4\cdot10^5~\rm{GeV}$ which is achieved after about $10^2R/c$ seconds of propagation, proton-proton interactions still occur, however at a rate of $\approx10^{-4}$ interactions per plasmoid radius out of which two orders of magnitude of reduction originate from geometric effects. In the simulation $pp$-interactions are observed to occur for distances as far as $10^3R$ at an average of $10^{-6}$ interactions per plasmoid radius. Beyond this distance protons would still undergo proton-proton interactions, yet it became statistically unlikely for the remaining protons to interact as interaction probabilities fell below the absolute amount of simulated particles. 

Decreases in energy to protons beyond from about $10^2R$ to $10^4R$ are basically dominated by proton synchrotron losses. The gyro-radii of the proton population which is proportional to $\propto E_p/B$ dropped from $0.3\,R$ to about $5\cdot10^{-2}R$ from the start of the simulation until the energetic $p\gamma$-plateau is reached due to the comparably high amount of energy being lost into the production of pions. Beyond that, the gyro-radius remains approximately constant as until the end of the simulation as the decrease in proton energy due to hadronic interactions and synchrotron losses are caused by the decrease in energy of the magnetic field due to geometric scaling along the jet. Hence, protons enter the diffuse regime after having interacted off most of their energy into pions in the first $10^{-1}R/c$ seconds which leads to an approximately constant rate of protons diffusing out of the plasmoid volume up to distances of $\approx 10^2\,R$. Protons remaining inside the plasmoid for more than $10^2\,R/c$ seconds are mostly confined to the plasmoid as implied by their comparably low gyro-radii in the order of $10^{-2}R$. 

\subsubsection{Neutrons}
\label{subsubsec:neutrons}

Photo-hadronic interactions of protons in most cases lead to a ``conversion'' of the proton into a neutron (yellow circles) and the emission of pions. Conversely, neutrons are ``re-converted'' into protons under further emission of pions upon interaction with another photon which is proportional to their likelihood of interaction with the photon backgrounds. During the first $0.5R/c$ seconds the energetic evolution of the neutron population is basically identical to the energies of protons inside the plasmoid as neutrons solely originate from protons and are re-converted into protons under emission of further pions. During this period, the numbers of protons and neutrons are approximately equal which, we have visualised in Fig.~\ref{fig:figure7_plasmoidImages} from actual simulation data, until also neutrons reach an energetic plateau where $p\gamma$-interactions become less likely. At these energies, neutrons also interact by two orders of magnitude less likely with the accretion disc field, in the order of one interaction per plasmoid radius. This leads to a neutron-induced energy leak of the plasmoid as neutrons are not deflected / confined by the magnetic field present and leave the plasmoid on a straight trajectory as also described in \citet{atoyan2003neutral_n1e1_neutralCurrents}. In the rest frame of the plasmoid, at average energies in the order of $10^6~\rm{GeV}$ in the plateau, neutrons would freely decay after a distance much larger than the plasmoid radius which can be estimated by
\begin{equation}
    L_{\rm{n,decay}}=\left<t_{n}\right>c\cdot\frac{E_n}{m_nc^2}\approx10^{17}~\rm{m}=10^4\,R\gg R
\end{equation}
with the mean neutron life time of $\left<t_{n}\right>\approx 900~\rm{s}$ and $E_n$ and $m_n$ as the neutron's energy and rest mass. As a result, there is a neutronic leak of energy due to neutrons escaping the plasmoid which with by two orders of magnitude higher than the rate of energy lost due to magnetically deflected protons escaping the interaction region. This effect is maximal after approximately $2-3R/c$ seconds as seen from the surface of the plasmoid and steadily decreases from this point on due to the decrease of interactions of nucleons with the accretion disc photon field. After all, the plasmoid will have lost roughly 90 per cent of its nucleonic primary particles in our simulation (protons and neutrons) after about $10R/c$ seconds of propagation.

\subsubsection{Neutrinos}
\label{subsubsec:neutrinos}

Neutrinos are produced in the decay of charged pions that emerge from $p\gamma$- (Eq.~(\ref{eq:processPGamma})) and $pp$-interactions (Eq.~(\ref{eq:processPP})). The neutrino population in our simulation can thus be divided into those being produced in photo-hadronic processes (green squares) and proton-proton interactions (green triangles) whereas the former case does not include any neutrino from interactions of a primary particle with the synchrotron photon field as contributions from this interaction channel were negligibly small.

\paragraph*{Neutrinos from $p\gamma$-interactions} During the first $0.5\,R/c$ seconds, nucleons reach the energetic plateau after having lost two orders of magnitude of average energy into interactions with the accretion disc field. The result of these interactions is a neutrino flare during this episode which extends until $p\gamma$-interactions become less efficient. As seen from an observer comoving on the surface of the plasmoid, this neutrino flare reaches a maximum not later than $t_{\max}\leq 2R/c$ seconds (vertical, dashed line) as at this point of time, the neutrino emission of each isotropically radiating volume element inside the plasmoid has reached each point on the plasmoid's surface. Indeed, the maximum number of observed neutrinos in the simulation lies at values exceeds the number of primary particles by more than one and almost two orders of magnitude at times around $\approx 1.5\,R/c$ seconds up to which the average energy of these neutrinos lies in the order of $\approx 10^6~\rm{GeV}=1~\rm{PeV}$. After the maximum of the flare, neutrinos generated by interactions with the accretion disc field steadily decrease in numbers until photo-hadronic interactions reach the lower threshold which is reached at about $10^2\,R/c$ seconds.

\paragraph*{Neutrinos from $pp$-interactions} Due to the cross section for proton-proton interactions being approximately constant over energies that primary particles achieved in the simulation (comp.\ Eq.~(\ref{eq:sigmaPP})), neutrinos are produced throughout the simulation in proton-proton-interactions. Hence, the maximum amount of neutrinos being produced in this interaction is also achieved after $\approx 2\,R/c$ seconds beyond which one should expect the number of neutrinos to be constant. However, the $r^{-2}$-decline in proton number density significantly reduces the production rate of neutrinos which beyond a distance of $10^3\,R$ results in no further neutrino being detected as $pp$-interactions have become statistically infeasible given the absolute amount of primary particles left in the simulation. Note that this holds true for an initial proton density of $n_{\rm{plasma},0}=10^{15}~\rm{m^{-3}}$ whereas the effects of lower and higher densities will be evaluated in the discussion section. Due to the considerably lower threshold energy of proton-proton-initiated neutrino production in the order of a few hundred $\rm{MeV}$ where pion production becomes possible, neutrinos emerging from this interactions persist to be emitted for a longer amount of time during the plasmoid's propagation. The average energy of $10^5~\rm{GeV}$ for neutrinos produced via this interaction channel tends to lie one order of magnitude below the average energy of neutrinos produced in $p\gamma$-interactions.

\subsubsection{Gamma-Rays}
\label{subsubsec:photons}

In analogy to neutrinos, gamma-rays are produced in this simulation as secondaries of $p\gamma$-interactions (red diamonds) or $pp$-processes (red crosses) yet are subject to absorption by photons of the accretion disc field and/or the electron synchrotron photon field and thereby producing $e^{+}e^{-}$ pairs. Without this effect, the observed gamma-ray emission of both the photo-hadronic and proton-proton interaction would exactly trace each the neutrinos' signals: In agreement with \citet{kelner2008protonGamma} for $p\gamma$-interactions, the simulation yields an average energy of gamma-rays that roughly lies half an order of magnitude above the average energy of neutrinos produced in this process yet with a ratio of absolute numbers of neutrinos to gamma-rays of approximately 2:1 due to an excess in charged pions. Similarly in $pp$-interactions, we reproduce the results of \citet{kelner2006protonProton} by finding that the average energies of photons and neutrinos produced in this process are roughly equal and that the numbers of neutrinos to gamma-rays produced also satisfies a ratio of approximately 2:1.

However, due to $\gamma\gamma$-absorption, the ratios among photons and neutrinos as mentioned above are not observed from outside of the plasmoid. For instance, the average energy of all detected gamma-rays which are secondaries of photo-hadronic interactions appears to trace the energetic signature of the proton and/or neutron populations and not the neutrinos' one. During the first $\approx 3\cdot 10^{-2}R/c$ seconds of propagation, an observer would see the average energy of photons non-dampened and following the signal of neutrinos produced in the same process. These photons, however, mostly originate from the outer regions of the plasmoid where $\gamma\gamma$-absorption is less significant due to the comparably short propagation length of about $10^{-2}R$ to escape the plasmoid. This effect yields a detected maximum of gamma-ray emission from photo-hadronic interactions at around $2\cdot 10^{-2}$ seconds beyond which the number of gamma-rays able to leave the plasmoid rapidly decreases. The more the plasmoid propagates, the more likely it becomes for a gamma-ray to be emitted closer from the center of the plasmoid as well as at lower energies as protons of high tend to diffuse out of the plasmoid. Hence, the longer the propagation length of a ensemble of photons, the more photons are absorbed on average by the background fields. However, since the optical depth of both the black body field and the electron synchrotron field are the most transparent to $\gamma\gamma$-absorption at roughly $10~\rm{TeV}$, the average energy of all observed gamma-rays will converge towards this value which one observes in our simulation after approximately $R/c$ seconds of propagation time. Low TeV-opacities in presence of a synchrotron field inside a plasmoid of comparable parameters have e.g.\ also been found in \citet{giannios2009fast_blobSize_blobProcesses}. 

\begin{figure*}
	\includegraphics[width=\textwidth]{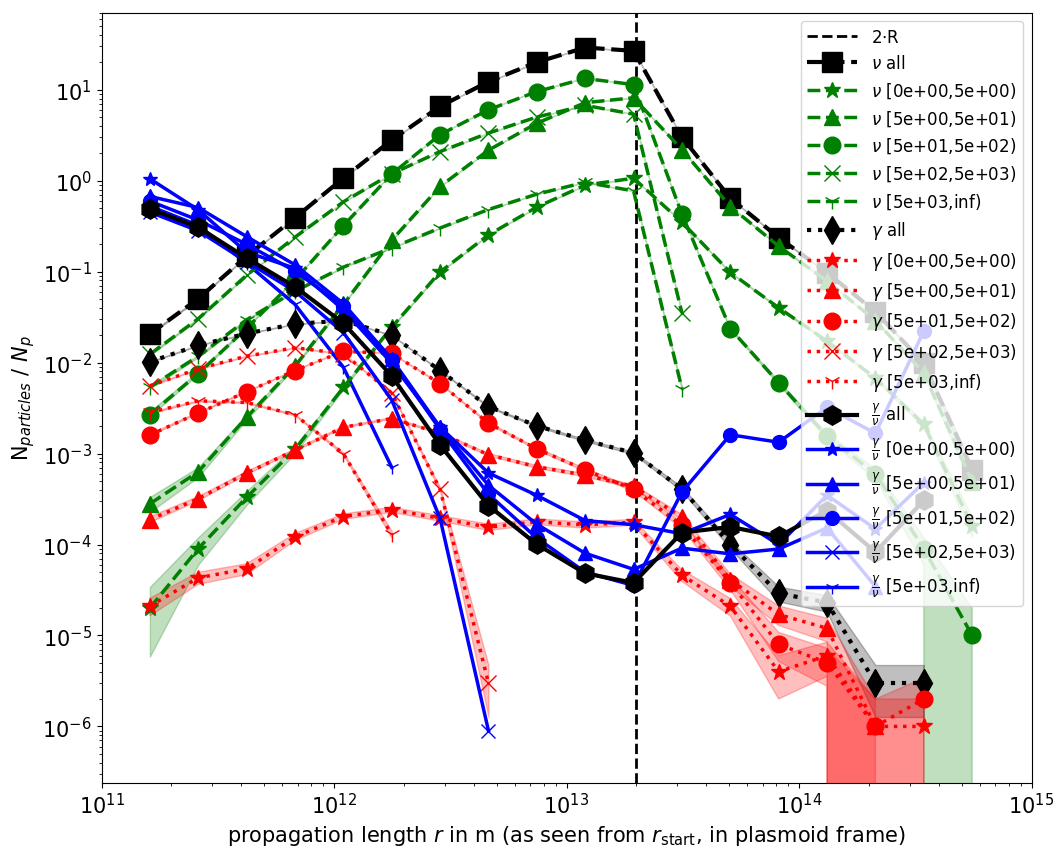}
	\caption{Ratios (solid lines) of the total number of neutrinos (dashed lines) and gamma-rays (dotted lines) over propagation length or time at selected energies in units of TeV. Propagation distances and energies are given in the plasmoid rest frame.}
	\label{fig:figure5_ratiosGammaNu}
\end{figure*}

The effect of $\gamma\gamma$-pair attenuation on the ratio of neutrinos to gamma-rays produced by $p\gamma$-interactions over propagation length / time is illustrated in Fig.~\ref{fig:figure5_ratiosGammaNu} at selected energy bins. Here, the black lines resemble the total neutrino (squares) and gamma-ray (diamonds) numbers relative to the amount of injected primary protons $N_p$ from Fig.~\ref{fig:figure4_plasmoidEvolution} (middle panel). The ratio of these curves $N_\gamma/N_\nu$ (black hexagons) yields a suppression of the total amount of gamma-rays to the total number of neutrinos of up to four orders of magnitude within the first $2R/c$ seconds due to the above mentioned production and absorption conditions. Beyond this time, when most of the remaining primaries' energies have been dissipated into the flaring event, the ratio remains roughly stable around values of $10^{-4}$ where the energies of primary protons has reached the plateau due to the less frequently occurring interactions of primaries with the accretion disc field below $\approx 10^6~\rm{GeV}$ (comp. Fig.~\ref{fig:figure3_opticalDepth}) where therefore also a plateau of the secondary particles' energies is reached (comp.\ Fig.~\ref{fig:figure4_plasmoidEvolution}, upper panel). Gamma-rays having escaped the plasmoid, thus not being tracked by our simulation anymore, further interact with the environment and thus even more alter the ratio of secondary gamma-rays to neutrinos. These gamma-rays most likely cascade down to lower energies as e.g. discussed in \citet{halzen2019neutrino_EMCascades} and would have to be taken into account for a fit to actual observational gamma-ray data of TXS 0506+056.\\

\begin{figure}
	\includegraphics[width=\columnwidth]{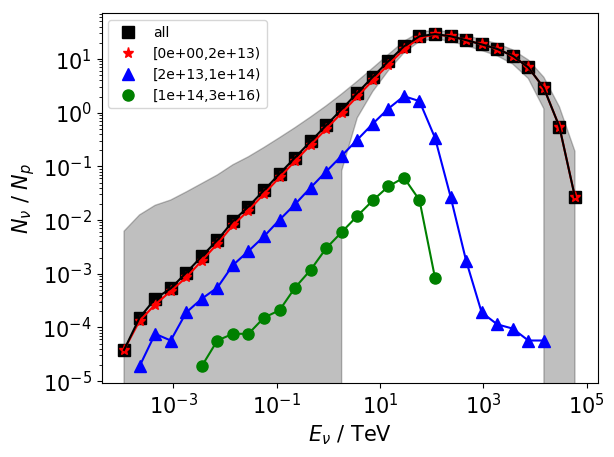}
	\caption{Spectra of neutrinos produced in the simulation. Energies and propagation distances given in plasmoid rest frame.}
	\label{fig:figure6_neutrinoSpectra}
\end{figure}

\begin{figure}
	\includegraphics[width=\columnwidth]{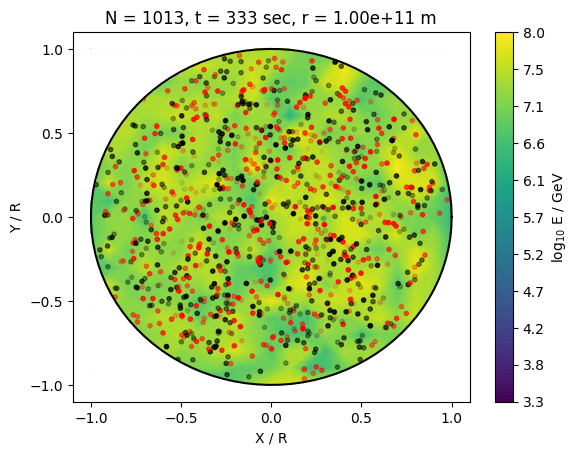}\\
	\includegraphics[width=\columnwidth]{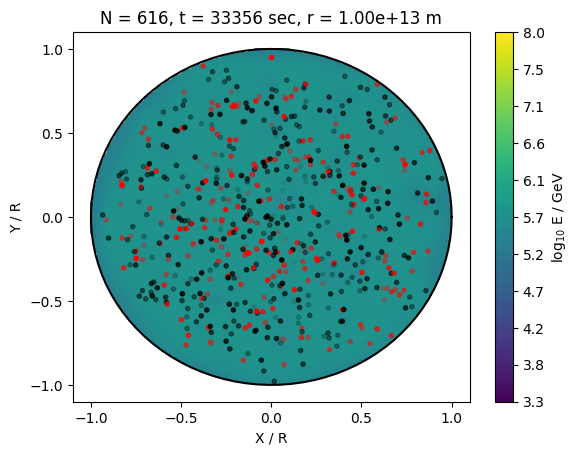}\\
	\includegraphics[width=\columnwidth]{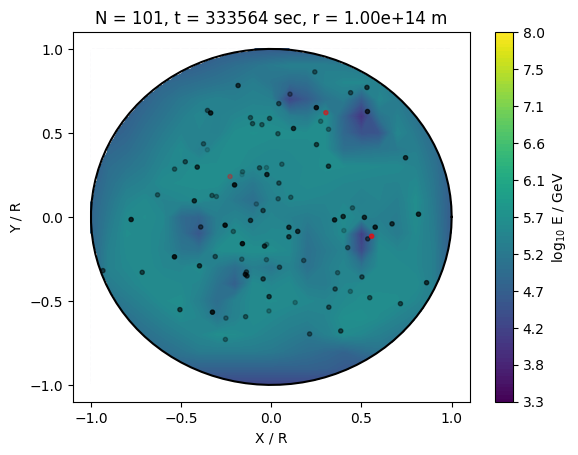}
	\caption{Visualisation of the plasmoid's evolution as seen by a comoving observer viewing from the jet axis using parameters as specified in Sec.~\ref{subsec:parameterSetup}. Black and red dots represent protons and neutrons, respectively whereas the dots' opacity indicates the positions of particles inside the plasmoid relative to the observer. The background colour represents the distribution of energy inside the plasmoid which throughout the simulation remains approximately homogeneous. All figures are stills from actual simulation data with a reduced total number of injected primaries for reasons of illustration.}
    \label{fig:figure7_plasmoidImages}
\end{figure}

The energetic evolution of all secondary neutrinos produced in $p\gamma$-interactions with the accretion disc black body field is shown in Fig.~\ref{fig:figure6_neutrinoSpectra} at selected propagation times. Black squares represent the spectrum of all neutrinos produced in the simulation of which the vast majority is produced within the first $2R/c$ seconds of propagation (red asterisks). Within this period, most neutrinos are produced at an energy of roughly $120\pm20~\rm{TeV}$ measured in the plasmoid frame.

\subsection{Impact of Parameter Variations}
\label{subsec:impactOfParameterVariations}

\subsubsection{Contribution of $pp$-Interactions}
\label{subsubsec:densityOfProtonPlasma}

One may interpret interactions among protons to either happen as interactions between primary protons only or as an interaction between one primary proton and another either less energetically significant or additionally occurring proton population. Subsequently, we will investigate either case and derive the implications of these scenarios on the evolution of our plasmoid.

If one interpreted all proton-proton-interactions to solely occur among primary, highly relativistic protons, equipartition arguments may be assumed to provide estimates on the energy densities of protons and electrons relative to magnetic fields. In this case, the initial magnetic field strength $B_0$ constrains the energetic boundaries of particles since the energy densities of protons and electrons scale $U_{p/e}\propto B_0^2$ as shown in Eq.~(\ref{eq:equipartition}). For the more general case of a relativistic electron-proton plasma we assume proton densities to be energetically power-law distributed with a spectral index $\alpha_p$ according to $\mathrm{d}n_p/\mathrm{d}E_p=\kappa_p(E_p/m_pc^2)^{-\alpha_p}$. We thus estimate the their number density by first calculating the proton energy density
\begin{equation}
    U_p = \kappa_p m_p^2c^4 \int_{\gamma_{p,\min}}^{\gamma_{p,\max}} \gamma_p^{1 - \alpha_p} \mathrm{d}\gamma_p
    \label{eq:calcProtonEnergyDensity}
\end{equation}
and the proton number density
\begin{equation}
    n_p = \kappa_p m_p c^2 \int_{\gamma_{p,\min}}^{\gamma_{p,\max}} \gamma_p^{-\alpha_p} \mathrm{d}\gamma_p
    \label{eq:calcProtonDensity}
\end{equation}
in which $\kappa_p$ is a normalisation constant with units $\rm{J}^{-1}~\rm{m}^{-3}$ and $\gamma_{p,\min/\max}$ are the minimal and maximal Lorentz factors of the proton population, one achieves the average energy of the proton population $\left<E_{p}\right>=U_p/n_p$ using the expressions in Eqs.~(\ref{eq:calcProtonEnergyDensity}) and (\ref{eq:calcProtonDensity}) above. If the energy densities of both protons and electrons are distributed in a relation $\chi=U_p/U_e$ to each other in the jet one may, using Eq.~(\ref{eq:energyBudget}), estimate the number density of protons via
\begin{equation}
    n_p = \frac{\chi}{\chi+1}\cdot \frac{U_b}{\left<E_{p}\right>}\approx 10^{-1}\frac{\chi}{\chi+1}\left(\frac{B_0}{1 \rm{G}}\right)^{2}\left(\frac{\left<E_{p}\right>}{10^8\,\rm{GeV}}\right)^{-1}\rm{m}^{-3}
\end{equation}
in which the average proton energy is mostly dominated by $\gamma_{\rm{p,\min}}$ for the case of a power-law proton spectrum. In our work where we have injected a monochromatic population of primary protons with $\left<E_p\right>=10^8~\rm{GeV}$ into a magnetic field of strength $B_0\approx1~\rm{G}$ one achieves, assuming that $\chi\approx 1/100$, a value for the average amount of interactions of a primary proton with another primary in the order of $\tau=n_p\sigma_{pp}R\approx 10^{-19}$ per plasmoid radius. Compared to photo-hadronic interactions with the accretion disc, which protons on average interact with about 100 times per plasmoid radius (above the threshold for this process), we can safely assume that energy-losses of primaries into $pp$-interactions are subdominant. Given the nature of our simulation which tracks individual particles in space and time during their propagation, we are not computationally able to inject a sufficient amount of particles to achieve a spectrum of secondary particles. In fact, the amount of particles we inject lies in the order of $10^6$ which on average would lead to one proton interacting within $10^{13}$ simulations of this kind. Hence, in order to achieve a feasible simulation efficiency, we up-scale the proton density up to an amount at which we achieve a sufficient amount of interactions to derive a spectrum of secondary particles while also ensuring that their contribution to the energy-loss of primary particles is comparably small. By this approach we are able to find a maximal proton plasma density at which we guarantee that losses into proton-proton interactions are always sub-dominant such that the evolution of the plasmoid is not significantly altered compared to a case in which $pp$-interactions would not occur as shown in Fig.~\ref{fig:figure3_opticalDepth}. We quantify this maximal plasma density by choosing $n_{0,\rm{plasma}}$ such that the average energy losses of $pp$-interactions compared to energy losses into $p\gamma$-interactions satisfy a ratio of $1/100$ when $p\gamma$-interactions are the most dominant process which we find to be in the first $\approx R/c$ seconds of the simulation until the energetic plateau is reached. During this episode we observe that the primary proton population losses two orders of magnitude of energy into interactions with the accretion disc field, i.e.\ $\mathrm{d}E_p/\mathrm{d}r|_{p\gamma}\approx 10^8~\rm{GeV}/R$. Hence, in order to achieve an energy loss ratio of roughly 1/100, the average energy lost into $pp$-interactions during the same propagation length would have to lie two orders of magnitude below this value, i.e.\ $\mathrm{d}E_p/\mathrm{d}r|_{pp}\approx 10^6~\rm{GeV}/R$. \citet{eichmann2012plasmoidHadronic} estimate the energy loss of primaries into proton-proton interactions per plasmoid radius to be
\begin{equation}
    \begin{split}
        &\left.\Dot{\gamma}_p\right|_{pp}=7\cdot10^{-6}\gamma_p\cdot\left(\frac{n_p}{10^{16}~\rm{m}^{-3}}\right)~\Rightarrow\quad\times\\
        &\quad\times\left.\frac{\mathrm{d}E_p}{\mathrm{d}r}(n_p)\right|_{pp}\approx 10^6\left(\frac{E_p}{10^8~\rm{GeV}}\right)\left(\frac{n_p}{10^{15}~\rm{m}^{-3}}\right)~\frac{\rm{GeV}}{R}
    \end{split}
    \label{eq:lossLengthPP}
\end{equation}
which for a proton population injected with an average energy of $E_p=10^8~\rm{GeV}$ requires a proton density in the order of $n_p\approx 10^{15}~\rm{m}^{-3}$ which is the value we have used in our simulation. 
The advantage of choosing this approach lies in the fact that the $pp$-neutrino and gamma-ray curves in Fig.~\ref{fig:figure4_plasmoidEvolution} may linearly be down-scaled in order to fit them to other models of lower ambient proton densities.

In literature jets are found rarely be dominated by proton-proton interactions as the proton densities required to generate significant amounts of neutrinos and gamma-rays would have to assume very high values as pointed out in \citet{aharonian2000tev_HadrIntnegligible_pgammaUpto10perCent, atoyan2003neutral_n1e1_neutralCurrents}. For instance, purely hadronic-dominated emission models tend to not explain several observations as pointed out in \citet{aharonian2000tev_HadrIntnegligible_pgammaUpto10perCent} who would need to infer densities in the order of $10^{12}~\rm{m}^{-3}$ in order to fit TeV-photon-flares at the example of Markarian 501. Considerably higher proton densities have been considered by \citet{eichmann2012plasmoidHadronic} in a pick-up model of protons in which a plasmoid absorbs matter from the jet during propagation as well as by \citet{celotti1998thermalMaterialJets} who review the possible amounts of non-relativistic, thermal protons in AGN jets. In another scenario \citet{gergely2009spin_flip} consider AGN-jets of former binary black hole systems to drill through dense environments after having experienced a re-orientation after the merger of their host black holes. Large proton densities could be achieved in scenarios in which a dense matter cloud enters the base of an AGN jet (\citet{dar1997hadronic_cloudEntersJet}) which may lie in the region of $10^{16}~\rm{m}^{-3}$ as modelled by \citet{araudo2010gamma_cloudEntersJet}. Densities of this order of magnitude do energetically contribute in the order of $1/10$ to the energy loss of highly relativistic, primary protons (comp.\ Eq.\  \ref{eq:lossLengthPP}) and would more significantly alter the evolution of a plasmoid with parameters as used in our work. Proton densities comparable to our estimate of $n_p\approx10^{-3}~\rm{m}^{-3}$ in Eq.~(\ref{subsubsec:densityOfProtonPlasma}) are also found by considerations done by \citet{reynolds1996matter_np1e-2_jetIsLeptonic, atoyan2003neutral_n1e1_neutralCurrents, dunn2006using_densitiesFit} who infer $pp$-interactions to solely occur among primary protons in AGN jets. In summary, we find the contribution of $pp$-interactions among primary protons to be negligible to the evolution of the plasmoid and therefore also to the absolute amount of secondary neutrinos and gamma-rays. Up to a proton density of $n_{0,\rm{plasma}}=10^{15}~\rm{m}^{-3}$, which we used in our simulation, the emission of secondary particles is still dominated by photo-hadronic interactions above the threshold for pion production while yielding sufficient statistical data to achieve a spectrum of secondary particles over time. Hence, for any proton density below our used value, Fig.~\ref{fig:figure4_plasmoidEvolution} overestimates the number of secondaries from $pp$-interactions by a factor of $n_p/(10^{15}~\rm{m}^{-3})$.

\subsubsection{Magnetic Field Strength \& Injection Energies}
\label{subsubsec:discussion_BandE}

The initial magnetic field strength $B_0$ significantly contributes to the evolution of the entire plasmoid system. For instance, particle energy densities scale proportional to $U_{p/e}\propto B_0^2$ in equipartition (comp.\ Eq.~(\ref{eq:equipartition})). For electrons this leads to a modified energy loss into synchrotron radiation whose photon number density in our model scales as $j_{\rm{synch}}\propto B_0^2\cdot B^{(\alpha_e+1)/2}(r)$. For magnetic fields stronger than one Gauss, in the region of $B\geq 10~\rm{G}$ as estimated in \citet{giannios2009fast_blobSize_blobProcesses}, the resulting synchrotron photon field would start to become the dominant target for photo-hadronic interactions compared to the accretion disc field and render the plasmoid even less transparent to high-energy gamma-rays. In such a case, secondaries produced in the $p\gamma$-process will start scaling according to the energy density of the magnetic field whereas the duration of the observed neutrino and gamma-ray flare will change $\propto U_b^{-1}$ due to the dependence of the energy loss length of protons on the number of target photons. Lastly, very high magnetic fields allow for higher proton densities to exist in energetic equipartition thus intensifying the flare in proportion to the magnetic field density. Smaller magnetic field strengths accordingly weaken the emitted synchrotron field. As in our simulation this would not have an effect on the low significance of $p\gamma$-interactions with this field, yet decreases the gamma-ray opacity of the plasmoid for photon energies beyond $E_\gamma>10^{10}~\rm{TeV}$ in proportion to the synchrotron emission coefficient $j_{\rm{synch}}\propto B^{(1+\alpha_e)/2}$. Due to this decrease we find the plasmoid's optical depth for gamma-ray absorption to basically be dominated by only the accretion disc field for magnetic field strengths of $B\leq 10^{-2}~\rm{G}$. The intensity of the observed flares of secondary particles would furthermore decrease due to fewer particles residing in the jet as inferred from energetic equipartition. The initial magnetic field of $B_0=1~\rm{G}$ which has also been used by \citet{inoue1996electron_B1g, eichmann2012plasmoidHadronic, christie2018radiative_EquilibriumPlasmoids_B1G, keivani2018multimessenger_blazarNuGamma} therefore corresponds to a case in which $p\gamma$-interactions are dominated by the accretion disc field whereas gamma-rays above TeV-scales tend to be absorbed by interactions with the synchrotron field. Note that our simulation terminates to track particles after escaping the spherical plasmoid volume. These particles would be subject to more (absorption) processes as the plasmoid is embedded into the AGN jet which is significantly longer than the radial extent of the plasmoid.\\

The amount of energy $E_{p,\rm{inj}}$ the primary protons are injected with determines the relevance of their interactions with the environment and hence the emission of secondary particles as summarised in Fig.~\ref{fig:figure3_opticalDepth}. Due to the dominant energy loss of protons with the accretion disc background, the emission scenario of the plasmoid can be divided in two cases:
\begin{enumerate}
    \item If the average energy of the primary population surpasses the threshold energy for $p\gamma$-interactions with the black body field at roughly $4\cdot 10^5~\rm{GeV}$, primary energy is rapidly converted into secondaries leading to a comparably intense neutrino flare. In this photo-hadronic scenario the flare duration is determined by the the amount of energy above the production threshold left to be converted into secondaries and the amount of time in which these particles reside in the plasmoid itself. Hence, the more energetic the primary population is the more intense becomes the resulting neutrino flare.

    \item For average energies below the $p\gamma$-threshold, proton-proton interactions dominate the production of secondary particles. In this hadronic scenario where the average energy of the proton population lies below roughly $10^5~\rm{GeV}$, energy losses are comparably small as shown in Eq.~(\ref{eq:lossLengthPP}). A plasmoid of our parameters would not produce a neutrino flare at these energies rather than continuously radiate its energy away into secondaries of $pp$-interactions and proton synchrotron radiation.
\end{enumerate} 

In our photo-hadronic setup we have chosen an injection energy of $E_{p,\rm{inj}}$. On the one hand, for injection energies significantly above this value, particles would still undergo a comparable amount of interactions per propagation length yet their gyro-radius would exceed the extent of the plasmoid and thus the vast amount of initial energy leaves the plasmoid as primary particles on almost approximately straight trajectories. The flare duration of such a scenario would be comparable to the situation presented in this work yet produce neutrinos of higher energies. On the other hand, if the injection energy lied significantly below $10^8\rm{GeV}$, particles would be more confined to the plasmoid region due to their decreased gyro-radius. However, neutrino emission would be affected in dependence on the amount of $p\gamma$-interactions with the black body photon field as implied in Fig.~\ref{fig:figure3_opticalDepth} and illustrated with the example of the following injection energies:
\begin{itemize}
    \item $E_{p,\rm{inj}}=10^7~\rm{GeV}$: The plasmoid evolution in this scenario would not significantly be different from our presented since the accretion disc field still is dominant at a comparable amount of interactions per propagation length. However, neutrino energies would be roughly an order of magnitude lower compared to Fig.~\ref{fig:figure4_plasmoidEvolution} and a reduced flare duration due to less energy being left in primaries for $p\gamma$-induced neutrino production.
    \item $E_{p,\rm{inj}}=10^6~\rm{GeV}$: Under our environmental parameters this injection energy corresponds to an edge-case towards a more hadronically dominated scenario where the average amount of interactions of primaries with the accretion disc field starts to decrease, leading to a shorter neutrino flare and eventually to one or less $p\gamma$-interactions per plasmoid radius and particle. In such an energetic setup the plasmoid suffers the highest amount of neutronic energy losses since primary protons are still being produced regularly yet the resulting neutrinos more rarely interact with the accretion disc field to yield a proton which would then be magnetically confined to the plasmoid again (comp.\ Sec.\ \ref{subsubsec:neutrons}).
    After this neutronic energy leak the plasmoid directly enters the more long-lived hadronic state in which $pp$-interactions and proton synchrotron radiation are the dominant radiation mechanisms.
    \item $E_{p,\rm{inj}}\leq10^5~\rm{GeV}$: A plasmoid of our parameters and injection energy would be entirely transparent to $p\gamma$-interactions and lack a neutrino flare comparable to Fig.~\ref{fig:figure4_plasmoidEvolution} as it would only produce neutrinos via $pp$-interactions. Since the gyro-radii of particles would be small compared to the plasmoid's size, particles were comparably well confined to the interaction region thus leading to no neutronic energy loss. The resulting larger amount of primaries remaining in the plasmoid then leads to an increased duration of $pp$-induced neutrino emission which correspondingly mostly depends on the number of interactions occurring a primary leaves the plasmoid.
\end{itemize}

\subsubsection{System Sizes and Proportions}
\label{subsubsec:discussionSizes}

The argumentation in Sec.~\ref{subsubsec:discussion_BandE} above neglects the finite extent of the plasmoid which imposes further constraints on the amount of energy that can be converted into secondaries and hence both the flare duration and intensity. Trivially, particles need to reside in the plasmoid on time-scales sufficient to interact with the environment which is governed by their gyro-radius $r_g$. Here, we refer to the gyro-radius per plasmoid radius $R=10^{13}~\rm{m}$ in multiples of the plasmoid radius which in terms of our choice of parameters can be expressed as
\begin{equation}
    r_g=3\cdot 10^{-1}\left(\frac{E_p}{10^8~\rm{GeV}}\right)\left(\frac{B}{1~\rm{G}}\right)^{-1}R
    \label{eq:gyroRadiusPerR}
\end{equation}
The evolution of the average gyro-radius of our plasmoid is illustrated in the lower panel of Fig.~\ref{fig:figure4_plasmoidEvolution} and is initially dominated by the comparably high energy loss of primary particles into $p\gamma$-interactions with the accretion disc field during the first $0.5\,R/c$ seconds of propagation. During this time, the gyro-radius drops from $r_g=0.3\,R$ below $10^{-2}R$ thus quickly entering the diffusive regime. For average proton energies at which photo-hadronic pion production is less significant or for distances further from the accretion disc a value for the gyro-radius of $r_g\geq 0.1\,R$ would not undergo such rapid declines and result in most of the primaries escaping the plasmoid beyond which point they are not modelled. Analogous arguments apply in which the magnetic field strength is smaller since $r_g\approx E/ceB$. Given a hypothetical value of $B_0=10^{-3}~\rm{G}$, in our simulation protons would have to have an average energy of $E_p \approx 10^5~\rm{GeV}$ in order to achieve the same, initial gyro-radius. However, this would lie below the threshold for $p\gamma$-interactions and hence the plasmoid would first experience a substantial leak in primary particles and afterwards radiate in the hadronic regime with its remnant population of primaries. Population of primaries with average energy above the photo-pion threshold in our scenario would have to be subject to stronger magnetic fields or exist in a plasmoid of larger radius to be effectively constrained to interactions inside the plasmoid volume.

The plasmoid size itself also constrains the lower value of positions from which the system is able to start propagating. In the presented model we have chosen a value of $r_{\rm{start}}=10^{14}~\rm{m}=10\,R$ to account for stronger magnetic fields and more intense photon fields. In particular with respect to interactions with the accretion disc black body field, this close distance yields comparably high energy losses in the first stage of the plasmoid's evolution resulting in rather more intense neutrino-flares as well as efficient primary particle containment within the plasmoid. Values of $r_{\rm{start}}<R$ from the central black hole are less likely due to geometric reasons whereas considerably larger values yield less efficient particle production as the occurrence of interactions are reduced by geometric effects (comp. Eqs.~(\ref{eq:BfieldScaling}), (\ref{eq:ambientMatterFieldScaling}), (\ref{eq:scalFuncSynchField}), (\ref{eq:scalFunctAccDisc}) as well as Fig.~\ref{fig:figure4_plasmoidEvolution}, lower panel). As a consequence, energy losses of primaries tend to be less efficient leading to stronger constraints on the initial proton energy and/or the plasmoid radius. In Eq.~(\ref{eq:accDiscIntensity}) we have argued that the radius of the accretion disc $R_{\rm{acc}}$ determines beyond which distances the decline in the number density of the accretion disc field scales $\propto r^{-2}$. Given our model a distance of $r_{\rm{start}}\approx 10R$ would not significantly have changed the interaction rate with the accretion disc field and hence the proton energy loss length. Furthermore, we have neglected any inner structure of the accretion disc or more sophisticated general-relativistic effects as e.g.\ considered in \citet{li2005blackbodyAccDisc} which would have led to more elaborated models of the number density of photons. We nevertheless assume that the hottest and hence most radiating region of the accretion disc will be the closest to the black hole, so results are less likely to be sensitive to $R_{\rm{acc}}$ itself.

\subsubsection{Disc Luminosity and the Broad Line Region}
\label{subsubsec:discussionLuminosity}

In our model we use a toy disc model which radiates homogeneously over its entire surface $A_{\rm{acc}}$ at a single temperature of $T_0$. The purpose of having chosen this non-realistic type of accretion disc lies in the fact, that we are able to derive the most important effects of the accretion disc on the evolution of the primary proton population already with this approach thus keeping the model as simple as possible. We also find that the effects of more realistic accretion disc models to only have a weak impact on the evolution of the plasmoid as long as certain parameter limits are concerned which we will quantify in the following.

One tempting yet misleading approach could be to linearly correlate the total disc luminosity $L_{\rm{disc}}$ with the amount of interactions a primary proton undergoes with the accretion disc field. This may be true if geometric effects are neglected yet in the case of our plasmoid such geometric effects can dominate. Consider an extreme case of an accretion disc as used in this work yet with an (almost) infinite radial extent. The total luminosity of such a disk would diverge yet the photon density, which is the key parameter influencing the primary population, at a point e.g.\ on the jet axis would be finite since far-out regions of the accretion disc contribute less by about a factor $|\vec{r}-\vec{r}'|^2$ (comp. Fig.~\ref{fig:figure2_discScalingSketch}) thus leading to convergent results if integrated over the entire accretion disc surface. Invoking the intensity integral from Eq.~(\ref{eq:accDiscIntensity}) yielding the geometric scaling function of the accretion disc field in Eq.~(\ref{eq:scalFunctAccDisc}) with $\lim\mathcal{S}_{bb}(R_{\rm{acc}}\rightarrow\infty)=1$ thus simply implies that the point on the jet axis, $R_{\rm{acc}}$, at which the disc becomes a point-like source is never reached thus resulting in an slowly decreasing, almost constant photon density at each point on the jet axis. In a more realistic scenario, however, most of the accretion disc's luminosity is likely to be emitted from the central disc region anyways thus one could in a firsst attempt mathematically reduce the effectively radiating, radial extent of the accretion down to some radius $R_{\rm{eff}}$. Assuming that the vast amount of luminosity is radiated within a region comparable to the radial extent of the plasmoid, this accumulates to a disc luminosity in the plasmoid frame of
\begin{equation}
    L_{\rm{disc}}^{\rm{(plasmoid)}}=5\cdot 10^{37}\left(\frac{R_{\rm{eff}}}{R}\right)\cdot\left(\frac{T}{T_0}\right)^4\cdot\left(\frac{\delta}{10}\right)^{-3} W~~.
\end{equation}
where $R$ and $T_0$ are as in Tab.~\ref{tab:parameters} and $T$ the temperature of the black body accretion disc field.

Further contributions to the photon density inside the plasmoid may come from the broad line region (BLR) of the AGN. Although that \citet{murase2014diffuse_BLRstartsAt1e17cm} estimate the luminosity of the BLR to be $L_{\rm{BLR}}\approx 0.1L_{\rm{disc}}$, the energy density of the BLR photon field $U_{\rm{BLR}}$ gets boosted by a factor of $\Gamma^2$ into the plasmoid frame. To achieve a conservative estimate on the BLR photon intensity we consider the entire BLR luminosity to be uniformly distributed on an infinitesimally thick shell of radius $R_{\rm{BLR}}\gg R_{\rm{start}}$. This approach offers the advantage to artificially increase the BLR intensity in the central regions of the AGN since the, as it would be in reality, volume elements of the BLR which are residing at distances larger than $R_{\rm{BLR}}$ are all moved to the same, closest distance towards the center down to that point at which the beginning if the BLR itself is defined. In spherical coordinates the resulting power density is thus
\begin{equation}
    \rho_{\rm{P,BLR}}=\frac{L_{\rm{BLR}}}{4\pi R_{\rm{BLR}}^2}\delta(r - R_{\rm{BLR}})
\end{equation}
which using the intensity integral in Eq.(\ref{eq:intensityIntegral}) yields an isotropic BLR intensity of
\begin{equation}
    I_{\rm{BLR}} = 
    \frac{L_{\rm{BLR}}}{4\pi R_{\rm{BLR}}^{2}} \cdot \frac{1}{\left(1 - r/R_{\rm{BLR}}\right)^2}~~.
    \label{eq:IBLR_comoving}
\end{equation}
Applying that $R_{\rm{BLR}}\gg R_{\rm{start}}$ and dividing by the speed of light then yields the energy density of the BLR which in return gets boosted by a factor of $\Gamma^2$ into the plasmoid frame, one gets
\begin{equation}
    U_{\rm{BLR}}^{\rm{(plasmoid)}}\approx\frac{L_{\rm{BLR}}}{4\pi c}\cdot\left(\frac{\Gamma}{R_{\rm{BLR}}}\right)^2~~.
    \label{eq:UBLR_comoving}
\end{equation}
which is the same estimate as used in the work of \citet[Eq.(8)]{murase2014diffuse_BLRstartsAt1e17cm} yet derived from strictly geometric and conservative arguments. One should note here, that only that fraction of this expression is boosted into the plasmoid frame which is within a region towards which the plasmoid is propagating thus further reducing the effective value of the expression above.
One may now achieve the photon number density by furthermore dividing Eq.(\ref{eq:UBLR_comoving}) by the Dopper-boosted energy of the average BLR photon energy which e.g.\ \citet{murase2014diffuse_BLRstartsAt1e17cm} model to assume values around approximately 10 to 40 eV. Finally, one may compare the significance of interactions of primary protons with boosted BLR photons compared to the de-boosted accretion disc black body field. Using the average black body photon energy of 7.6 eV in the plasmoid frame and a value for $R_{\rm{BLR}}=1~\rm{pc}$ we achieve
\begin{equation}
    n_{\gamma,\rm{disc}}~/~n_{\gamma,\rm{BLR}} \approx 100
\end{equation}
as long as the plasmoid propagates within a distance of $r\ll R_{\rm{BLR}}$. This relation is likely to change over distance when approaching the BLR such that, given our derivation, the spatial dependence on $r$ in Eq.~(\ref{eq:IBLR_comoving}) would have to be applied. However, this in return is the result of our toy-BLR with the conservative estimate of the entire BLR luminosity being distributed over a shell of radius $R_{\rm{BLR}}$. A more realistic estimate of the scaling of the BLR photon density over distance could therefore be achieved by applying a modified expression of the luminosity density of the BLR to the geometric integral in Eq.~(\ref{eq:intensityIntegral}). Also noting that we have over-estimated the BLR contribution by invoking the entire BLR luminosity and not only the part towards which the plasmoid is propagating, one finds that in our model energy losses of primaries into interactions with BLR photons are not a dominant process. However, one should note that for larger distances from the black hole, the accretion disc field will be much weaker, eventually leading to the BLR photon field, yet also the synchrotron field, to become the dominant photon targets as also pointed out in \citet{atoyan2001high_BLRweaInBLLac, murase2014diffuse_BLRstartsAt1e17cm}.

\section{Summary and Conclusions}
\label{sec:conclusions}

We have presented the evolution of a primary population of high-energy protons residing in a plasmoid with radius $R=10^{13}~\rm{m}$ which propagates with a Lorentz factor of $\Gamma=10$ along an AGN jet for a distance of $10~\rm{pc}$ starting from a position $r_{\rm{start}}=10^{14}~\rm{m}$ as seen from the central black hole. The plasmoid was modelled to consist of an electron-proton plasma which is in energetic equipartition with the surrounding magnetic field (Eq.~(\ref{eq:equipartition})). The energy densities of protons and electrons are modelled to be distributed in a ratio of $\chi=U_p/U_e=1/100$. We assumed the magnetic field to be purely turbulent with an initial RMS field strength of $B_0=1~\rm{G}$ and took into account the presence of an accretion disc black body photon field of temperature $T_{\rm{black body}}=10~\rm{eV}/k_b$ (Eq.~(\ref{eq:photonDensityBlackbodyDeBoost})) as well as the synchrotron photon field generated by the plasma's relativistic electrons (Eq.~(\ref{eq:numberDensitySynchrotron})). We account for $p\gamma$-interactions, synchrotron radiation as well as inelastic $pp$-collisions of primary protons with either themselves or an energetically subdominant proton plasma present in the jet (Eq.~(\ref{eq:ambientMatterFieldScaling})). Scalings due to either energetic or geometrical considerations of the magnetic, photonic and hadronic fields have been included in the model. Gamma-rays, which all occur as secondary particles from the interactions above undergo $\gamma\gamma$ pair attenuation from photons of either the accretion disc or electron synchrotron field. All parameters used in our model can be found in Tab.~\ref{tab:parameters}.

We find for our simulation of an isotropic, homogeneous, and instantaneous injection of protons with energies $E_{\rm{start}}=10^8~\rm{GeV}$ a PeV neutrino-flare to occur which in the plasmoid frame peaks slightly before $2R/c$ seconds of propagation time and which is preceded by a fainter gamma-ray flare found to peak at roughly $10^{-1}\,R/c$ seconds. The neutrino flare is dominated by production via photo-hadronic interactions with the accretion disc field whereas a relative deficit of high-energy gamma-rays compared to neutrinos of up to a factor of $10^{-4}$ is caused by $\gamma\gamma$ absorption with either the black body field and/or the synchrotron field whose combined optical depths achieve the lowest values for gamma-rays in the TeV-regime. Hence, this model, in which particles are accelerated in plasmoids that are launched at the foot of the AGN jet, could solve the general question in high-energy neutrino astronomy why observations point toward a scenario in which the emission of gamma-rays must be suppressed:
\begin{enumerate}
    \item The first concerns the detection of the diffuse neutrino flux $\Phi_\nu$ as for a spectrum significantly steeper than $\Phi_\nu\propto E_\nu^{-2}$, the diffuse, extra-galactic gamma-ray background of Fermi would clearly be violated by the accompanying photons. Here, it may still be possible that $\gamma\gamma$ interactions lead to a cascade down to MeV energies as discussed in \citet{halzen2019neutrino_lessGammaThanNeutrino}. 
    \item The second concerns the evidence for a neutrino flare of the source TXS 0506+056 in 2014/2015 that lasted around 100 days and has a significance of $\sim 3\sigma$ of being an astrophysical neutrino flare from the direction of TXS 0506+056 \citep{icecube_txs2018}. This detection - as opposed to the flare in 2018 \citep{icecube2018multimessenger_PKS_NEHuGamma} - lacks a flare at gamma-ray energies and production close to the jet base would be a solution even here. However, that flare lasted about 6 months, so the injection scenario needs to be different from what is discussed here. Our flaring time-scales are much shorter due to the delta-type injection in time. Consequently, one could add a further layer of complexity to the model by choosing a Heaviside-type of injection such that there is a constant injection of primary protons into the plasmoid volume over some time interval $\Delta T$ enabling a fit to e.g.\ the flaring event observed in the context of TXS 0506+056. The length of that time interval would then depend on the injection luminosity which also in an even further step could be a function of time itself.
\end{enumerate}
In addition, the evolution of the primary protons and the flux of produced neutrinos is most likely not dominated by proton-proton interactions as for the case of hadronic neutrino production solely occurring among primary particles, the amount of produced neutrinos would be negligible compared to interactions of protons with the photon fields. However, less energetically significant proton targets of higher densities may indeed lead to higher contributions to the amount of produced neutrinos and for densities higher than $n_{\rm{plasma}}\approx 10^{16}~\rm{m}^{-3}$ even become the dominant neutrino production process.\\

Conclusively, our model parameters are compatible to the TXS 0506+056 detections yet the full modelling of the time evolution of the TXS SED goes beyond the scope of this paper and will be subject to future work. The model developed here is meant to be a first step in using our modified version of {\small CRPROPA} aiming to represent a flexible, numerical code including the relevant interactions and target environments that can be applied to very different emission scenarios in the future, thus opening a window to systematically investigating neutrino emission scenarios from AGN.

\section*{Acknowledgements}

We wish to warmly thank the German Academic Scholarship Foundation (Studienstiftung des deutschen Volkes) for funding the corresponding author and this work. We also thank Bj\"orn Eichmann for useful comments and Lukas Merten for comments on the structure of {\small CRPROPA} as well as Julia Ebeling for a prototype version of a {\small CRPROPA} hadronic interaction module. PJM would like to thank the Department of Physics at the University of Oxford. GC acknowledges support from STFC grants ST/S002952/1, ST/S002618/1, ST/N000919/1 and from Exeter College, Oxford.




\bibliographystyle{mnras}
\bibliography{literature} 


\label{lastpage}
\end{document}